\documentclass[AMA,Times1COL]{WileyNJDv5} 

\usepackage{threeparttable}
\usepackage{booktabs}
\usepackage{graphicx}

\newcommand{\logit}{\text{logit}}

\newcommand{\sigmai}{\sigma_i}

\newcommand{\sumi}{\sum_{i=1}^{N}}

\newcommand{\prodi}{\prod_{i=1}^{N}}

\newcommand{\p}{p}

\newcommand{\intR}{\int_{-\infty}^{\infty}}

\newcommand{\with}{\text{ with }}

\newcommand{\phidens}{\dfrac{1}{\tau}\phi\left(\dfrac{\thetai-\theta}{\tau}\right)}

\newcommand{\Yi}{y_i}
\newcommand{\Ni}{n_i}
\newcommand{\thetai}{\theta_i}

\newcommand{\si}{s_i}
\newcommand{\Ya}{y_{i0}}
\newcommand{\Na}{n_{i0}}
\newcommand{\Yb}{y_{i1}}
\newcommand{\Nb}{n_{i1}}
\newcommand{\smat}[4]{\begin{bmatrix} {{#1}} & {{#2}} \\ {{#3}} & {{#4}} \end{bmatrix}}

\newcommand{\Pmin}{P_\text{min}}
\newcommand{\Pmax}{P_\text{max}}
\newcommand{\nmin}{n_\text{min}}
\newcommand{\nmax}{n_\text{max}}

\articletype{}%

\received{26 April 2016}
\revised{6 June 2016}
\accepted{6 June 2016}

\begin{document}

\title{Copas-Heckman-type sensitivity analysis for publication bias in rare-event meta-analysis under generalized linear mixed models}

\author[1,2,3]{Yi Zhou}
\author[3,4]{Taojun Hu}
\author[1]{Yuji Sakamoto}
\author[5]{Ao Huang}
\author[2,4]{Xiao-Hua Zhou*}
\author[3,6]{Satoshi Hattori*}

\authormark{Zhou \textsc{et al.}}
\address[1]{\orgdiv{Division of Mathematics and Informatics, Graduate School of Human Development and Environment}, \orgname{Kobe University}, \orgaddress{\state{Kobe}, \country{Japan}}}
\address[2]{\orgdiv{Beijing International Center for Mathematical Research}, \orgname{Peking University}, \orgaddress{\state{Beijing}, \country{China}}}
\address[3]{\orgdiv{Department of Biomedical Statistics, Graduate School of Medicine}, \orgname{The University of Osaka}, \orgaddress{\state{Osaka}, \country{Japan}}}
\address[4]{\orgdiv{Department of Biostatistics}, \orgname{Peking University}, \orgaddress{\state{Beijing}, \country{China}}}
\address[5]{\orgdiv{Department of Medical Statistics}, \orgname{University Medical Center G{\"o}ttingen}, \orgaddress{\state{G{\"o}ttingen}, \country{Germany}}}
\address[6]{\orgdiv{Integrated Frontier Research for Medical Science Division, Institute for Open and Transdisciplinary Research Initiatives}, \orgname{The University of Osaka}, \orgaddress{\state{Osaka}, \country{Japan}}}

\corres{
Satoshi Hattori, Department of Biomedical Statistics, Graduate School of Medicine, The University of Osaka, Osaka, Japan.
\email{hattoris@biostat.med.osaka-u.ac.jp}\\
Xiao-Hua Zhou, Beijing International Center for Mathematical Research, Peking University, Beijing, China.
\email{azhou@math.pku.edu.cn}}

\abstract[Summary]
{
In systematic reviews and meta-analyses, publication bias (PB) is one of the serious concerns and mainly induced by selective publication of academic literatures. 
Although many methods have been proposed to deal with PB, almost all the methods are based on the normal-normal (NN) random-effects model assuming that data are normally distributed in both the within-study and the between-study levels.
For rare-event meta-analysis where data contain rare occurrences of events, the standard NN random-effects model may perform poorly.
Instead, some generalized linear mixed models (GLMMs) which employ the exact distribution for the number of events in within-study level provide alternatives and have been widely used in practice.
However, limited methods can be applied to deal with PB in the GLMMs.
To address this limitation, we propose a framework of sensitivity analysis for evaluating the impact of PB in various GLMMs. The proposed framework is developed based on the famous Copas-Heckman-type sensitivity analysis methods and can be easily implemented with the standard software with small computational cost.
In this paper, we conduct simulation studies to assess the performance of proposed methods in adjusting PB and compare the results with related existing methods.
Several real-world examples are also analyzed to show the broad applicability of our proposal in evaluating the potential impact of PB in meta-analysis of odds ratios and proportions with rare-event outcomes.
}


\jnlcitation{\cname{%
\author{Y. Zhou}, 
\author{T. Hu}, 
\author{A. Huang},
\author{Y. Sakamoto},
\author{X-H. Zhou}, and 
\author{S. Hattori}} (\cyear{2025}), 
\ctitle{Copas-Heckman-type sensitivity analysis for publication bias in rare-event meta-analysis under generalized linear mixed models}, 
\cjournal{xx}, \cvol{2025;00:1--6}.}

\maketitle

\section{Introduction}
\label{sec1}

Systematic reviews and meta-analyses are widely used in a variety of areas to summarize the overall effect of an intervention or the difference between two groups. 
Usually, the overall effect is estimated by a two-level random-effects model.
At the within-study level, it is assumed that the observed outcomes of each study follow a normal distribution with a study-specific true mean and a known variance of outcomes.
At the between-study level, the true means across studies are distributed as a normal distribution with an overall mean and a between-study variance, which are the parameters of interest.
This method is considered to be the standard approach for meta-analysis and is also known as the inverse-variance method,\cite{DerSimonian1986} the linear mixed-effects model, or the normal-normal (NN) random-effects model (hereinafter referred to as the NN model).
When the outcomes are binary, a common practice is to transform the outcomes into continuous values and apply a normal approximation in order to use the NN model.
For example, in a meta-analysis comparing two groups, it is often of interest whether an event (e.g., an adverse event) is more or less likely to occur in the treatment group.
The treatment effect is commonly quantified by the odds ratio (OR) or the log-transformed OR (lnOR), which compares the odds of the event between two groups.

A major issue with these measures arises when the observed number of events are close to or contains zero, resulting in the estimates of OR and its standard error (SE) being either zero or undefined.
To deal with this, continuity correction methods\cite{Sweeting2004} are commonly applied.
However, these methods are often criticized for lacking a sound theoretical rationale, and no single method is consistently recommended.\cite{Zabriskie2023}
Despite the limitations of the continuity correction methods, the small number of events, also referred to as rare events, can lead to problematic estimation when using the NN model.
In the example above, the lnORs are typically modeled using the NN model.
However, in situations of rare events or small sample sizes, the normal approximation used at the within-study level becomes questionable.
Furthermore, the NN model assumes independence between lnOR estimates and their standard errors (SEs), which may not hold in the presence of rare events and also introduce bias.\cite{Stijnen2010,Jansen2023}
Alternatively, generalized linear mixed models (GLMMs) offer a more accurate estimation by modeling the number of events using exact likelihoods for discrete random variables.\cite{Stijnen2010,Jansen2023,Hamza2008}
The advantages and performance of various GLMMs have been extensively evaluated through simulation studies.\cite{Jansen2023,Stijnen2010,Hamza2008,Cai2010,Kuss2015,Jackson2018,Mathes2018,Mathes2021} 
The widely used ones are the GLMM using hypergeometric or binomial distribution at the within-study level and a normal distribution at the between-study level;\cite{Hamza2008,Stijnen2010,Jansen2023} in this paper, these models are in general referred to as the hypergeometric-normal (HN) random-effects model (hereinafter referred to as the HN model) and the binomial-normal (BN) random-effects model (hereinafter referred to as the BN model), respectively.

In systematic reviews, a common and unavoidable issue is publication bias (PB)--a phenomenon in which studies with significant results are more likely to be published--potentially leading to over-optimistic conclusions in meta-analyses.\cite{Sutton2008}
In recent decades, many sophisticated methods incorporating selection functions into estimation frameworks have been proposed to quantitatively assess the potential impact of PB.\cite{Sutton2008,Jin2015}
Famous ones include the sensitivity analysis methods proposed by Copas and his colleagues, such as the Copas-Heckman-type sensitivity analysis methods,\cite{Copas1999,Copas2000,Copas2001} which extended the Heckman model\cite{Heckman1976,Heckman1979} in the field of econometrics into meta-analysis models.
Others include the $t$-statistic-based selection model\cite{Copas2013} and the non-parametric worst-case bounds.\cite{Copas2004}
Among these, the Copas-Heckman-type sensitivity analysis methods\cite{Copas1999,Copas2000,Copas2001} has been extensively developed\cite{Ning2017,Huang2021,Li2022,Huang2023} and further extended to address PB in multivariate meta-analyses.\cite{Hattori2018,Piao2019,Li2021,Mavridis2013}
Almost all of these methods were developed based on the NN model or its multivariate versions.
An exception is the method of Hattori and Zhou,\cite{Hattori2018} which extended the Copas-Heckman-type sensitivity analysis methods into the meta-analysis of diagnostic studies with bivariate exact within-study likelihood.

Although numerous exact likelihood-based models have been proposed for summarizing overall effects for rare-event meta-analyses, these models do not account for PB. 
To the best of our knowledge, limited research have been studied to address PB in models other than the NN model; the only known proposal is that of Hu et al.\cite{Hu2024}
This method extended the $t$-statistic-based selection model\cite{Copas2013} into the HN and BN models; it assumes selective publication to be related with the $t$-statistics (equivalently, $p$-values) of studies and successfully adjusts PB using a sensitivity analysis approach.
However, one of the limitations of this method is its high computational burden.
To facilitate estimation, its algorithm employs an approximation technique, which may also result in some degree of estimation error.\cite{Hu2024}
Moreover, the continuity correction is still required in constructing the selection function; thus, the estimations might be influenced by various approaches of continuity correction.
In this paper, we propose an alternative sensitivity analysis framework for addressing PB that is substantially simpler in terms of computational implementation.
Our proposal considered the selection functions related with the sizes of studies, thereby completely avoiding the need for continuity correction during estimation.
While PB is often attributed to statistical significance, it remains important to consider other potential sources, such as the size of studies, as part of sensitivity analysis to derive robust conclusions in meta-analysis.
Our proposal offers a practical alternative for addressing PB in the context of rare-event meta-analyses.

The rest of this paper is organized as follows.
In Section \ref{sec:copas}, we review the standard normal-normal random-effects meta-analysis model for summarizing the binary outcomes and present two Copas-Heckman-type sensitivity analysis methods.
In Section \ref{sec:glmm}, we introduce the commonly used GLMMs for rare-event meta-analyses. 
In Section \ref{sec:sa}, we propose the framework of novel sensitivity analysis methods for PB in different GLMMs.
The performances of the proposed methods in adjusting PB are studied using stimulation experiments in Section \ref{sec:simu}.
In Section \ref{sec:app} we demonstrate the practical applicability of the proposed methods through several real-world meta-analyses.
Finally, we conclude with a discussion in Section \ref{sec:diss}.
Some results are presented in Supplementary Material.

\section{Copas-Heckman-type sensitivity analysis for publication bias}\label{sec:copas}

\subsection{Normal-normal random-effects model}\label{sec2.1}

In this section, we focuses on meta-analysis under the assumption that PB is absent.
Suppose that we are interested in a meta-analysis containing $N$ studies with binary outcomes, where the studies are randomly published from the population of all relevant studies.
Each study contains treatment and control groups, and the numbers of events and subjects are accessible from clinical literature.
Let $\Ya$ and $\Na$ be the number of events and subjects in the control group, respectively, and $\Yb$ and $\Nb$ are those in the treatment group.
These data in study $i$ can be formulated by the contingency matrix in Table \ref{tab:conf}.

With two groups of data, the overall OR or its log-transformation (i.e., lnOR) is widely used as the effect measure of interest in clinical and epidemiological research. 
The common practice of meta-analysis is based on the NN model.\cite{DerSimonian1986}
For the $i$th study in Table \ref{tab:conf}, one can estimate the lnOR, denoted by $\hat\thetai$, and its SE, denoted by $\si$, with
\begin{align*}
\hat\thetai = \log \dfrac{\Yb/(\Nb-\Yb)}{\Ya/(\Na-\Ya)}
\text{~and~}
\si = \sqrt{\dfrac{1}{\Yb}+\dfrac{1}{\Nb-\Yb}+\dfrac{1}{\Ya}+\dfrac{1}{\Na-\Ya}},
\end{align*}
where the lnOR measures how the odds of the event of interest differ between the treatment and control groups.
{When zero entries appear in Table \ref{tab:conf}, the estimates of $\hat\thetai$ and $\si$ become undefined.
In such case, the continuity correction is usually conducted. 
There are many approaches to implementing continuity corrections, which involve adding different constants to some or all studies; however, these choices have some influence on the estimate of the overall lnOR.
Zabriskie et al.\cite{Zabriskie2023} extensively compared various approaches of continuity correction by a simulation study and found that no one-method could be recommended; Sweeting et al.\cite{Sweeting2004} suggested a sensitivity analysis across a number of correction constants.
A common approach is still to add 0.5 to both the numbers of events and non-events prior to estimating the lnOR and SE.\cite{Sweeting2004}} 

With the estimates of $\hat\thetai$ and $\si$, the NN model can be applied.\cite{DerSimonian1986} At the within-study level, it is assumed that the estimated lnOR approximately follows a normal distribution: 
\begin{align}
\hat\thetai \mid \thetai \sim N(\thetai, \si^2),\label{nn1}
\end{align}
where $\thetai$ is the true lnOR for each study; following the convention in meta-analysis field, $\si$ is regarded as known and estimated.
The normal approximation in the within-study model \eqref{nn1} holds when the sizes of studies are sufficiently large and the event probabilities are not extremely close to 0 or 1. 
At the between-study level, the true lnORs across the studies are assumed to have the following normal distribution:
\begin{align}
\thetai \sim N(\theta, \tau^2),\label{nn2}
\end{align}
where $\theta$ is the overall lnOR of interest, and $\tau^2$ is the between-study variance.
Marginally, by combining models \eqref{nn1} and \eqref{nn2}, $\hat\thetai$ has the following normal distribution:
\begin{align}\label{rem}
\hat\thetai \sim N(\theta, \si^2+\tau^2).
\end{align}
Allowing for the existence of between-study heterogeneity, the parameters $(\theta,\tau)$ can be estimated by the maximum likelihood (ML) method based on the following likelihood:
\begin{align}\label{llk1}
L(\theta,\tau)=\prodi \intR L_i(\thetai)\phidens~d\thetai
\with
L_i(\thetai) = \dfrac{1}{\sqrt{2\pi}\si}\exp\left(-\dfrac{(\hat\thetai-\thetai)^2}{2\si^2}\right),
\end{align}
where $\phi(.)$ denotes the probability density function of the standard normal distribution.

Sometimes, the occurrence of an event in only one group is of primary interest.
In such case, the parameter of interest is the event proportion, denoted by $\pi$. 
To take an example, $\pi$ can be the overall probability of an adverse event occurring in the treatment group. 
The standard approach is to model the log odds of the event, or equivalently, the logit-transformed proportion, denoted by $\thetai=\logit(\pi_i)$, where $\pi_i$ represents the study-specific true occurrence probability of event, and $\logit(x)=\log(x/(1-x))$ is the logit function.
The log odds can be estimated as $\hat\thetai$ based on data of the treatment group only in Table \ref{tab:conf}.
With the observed number of events ($\Yb$) and total number of subjects in the treatment group ($\Nb$), the log odds and its SE are estimated as 
\begin{align*}
\hat\thetai = \log \dfrac{\Yb}{\Nb-\Yb}
\text{~and~}
\si = \sqrt{\dfrac{1}{\Yb}+\dfrac{1}{\Nb-\Yb}}.
\end{align*}
{Similarly, the continuity correction is also required when there are zero events although it may cause some variants in the results.}
Under the large sample conditions, the estimated log odds can also be assumed to follow the normal distribution of model \eqref{rem}, allowing the overall effect parameter $\theta$ to be estimated using the ML method.
Finally, the parameter of interest, $\pi$, is obtained by transforming the estimated $\theta$.

\subsection{Copas-Heckman-type sensitivity analyses}\label{sub:copas_heckman_type_sensitivity_analysis}

PB has always been one of the major issues in meta-analysis.
In the presence of PB, the published $N$ studies are not likely to be random sample from the population. 
Copas and Shi\cite{Copas2000} proposed one sensitivity analysis method for addressing PB in the NN model \eqref{rem} by developing the Heckman-type selection function.\cite{Heckman1976,Heckman1979}

In this method, outcomes $\hat\thetai$ are re-expressed as follows:
\begin{align*}
\hat\thetai = \theta+{(\sigma_i^2+\tau^2)}^{1/2}\epsilon_i,  
\end{align*}
where $\epsilon_i$ is the standard normal residual, and $\sigma_i^2$ is the within-study sampling variance, which can be replaced by $\si^2$ when numbers of subjects in each study are large.
Copas and Shi\cite{Copas2000} conjectured that selective publication process is related with the SEs of studies; thus a latent Gaussian variable $Z_i$ with the selection equation concerning the SE of each study was proposed, that is,
\begin{align*}
Z_i = \beta_0+\dfrac{\beta_1}{s_i}+\delta_i,
\end{align*}
where $\beta_0$ and $\beta_1$ are constants, and $\delta_i$ is the residual following the standard normal distribution, $N(0,1)$.
Study $i$ is published if and only if $Z_i>0$.
The correlation between $\epsilon_i$ and $\delta_i$, denoted by $\rho$, was introduced to link selective publication process with the outcomes.
Thus, the joint distribution of $\epsilon_i$ and $\delta_i$ can be written as:
\begin{align*}
\binom{\epsilon_i}{\delta_i}&\sim N\left( \binom{0}{0}, \smat{1}{\rho}{\rho}{1}\right).
\end{align*}
When $\rho\neq0$, the publication of a study is related with its outcomes; in contrast, $\rho=0$ indicates that the study outcomes do not influence its publication.
The probability of $Z_i>0$ can be re-expressed by the probit mode:
\begin{align*}
P(Z_i>0\mid \si) = \Phi\left(\beta_0+\dfrac{\beta_1}{s_i}\right),
\end{align*}
where $\Phi(.)$ is the standard normal cumulative distribution function. 
The probability of $Z_i>0$ represents selective publication of one study, indicating that studies with smaller SEs, which are usually large studies, are more likely to be published.

Under selective publication of studies, the log-likelihood conditional on the published studies was derived as follows:
\begin{align}\label{llkcs}
\ell(\theta,\tau,\rho,\gamma_0,\gamma_1)&=\sumi \log p(\hat\thetai\mid z_i>0, \si)\nonumber \\
&= \sumi\left\{ \log p(\hat\thetai)+\log p(z_i>0\mid \hat\thetai, \si) - \log p(z_i>0\mid\si)\right\} \nonumber\\
&= \sumi\left\{-\dfrac{1}{2}\log(\tau^2+\sigmai^2) - \dfrac{\hat\thetai-\theta}{2(\tau^2+\sigmai^2)} - \log\Phi(\gamma_0+\gamma_1/s_i) + \log\Phi(v_i) \right\}
\end{align}
with
\begin{align*}
v_i = \dfrac{\gamma_0+\gamma_1/s_i + \tilde\rho_i(\hat\thetai-\theta)/\sqrt{\tau^2+\sigmai^2}}{\sqrt{1-\rho_i^2}}
\text{ and }
\tilde\rho_i = \dfrac{\sigmai}{\sqrt{\tau^2+\sigmai^2}}\rho.
\end{align*}
This log-likelihood remains consistent with the NN model in the absence of PB (when $\rho=0$).
Since the likelihood for estimating parameters $\beta_0$ and $\beta_1$ tends to plateau,
Copas and Shi\cite{Copas2000} proposed a sensitivity analysis approach by fixing various pairs of values for $(\beta_0,\beta_1)$.
Fixing these values sets the selection probabilities $P(Z_i>0\mid \si)$, which then allows the remaining model parameters $(\theta, \tau, \rho)$ in the log-likelihood \eqref{llkcs} to be estimated via the ML method.
In addition, the number of unpublished studies can be be approximated as 
\begin{align*}
M=\sumi\dfrac{1-P(Z_i>0\mid\si)}{P(Z_i>0\mid \si)}.
\end{align*}
Since the true values of $(\beta_0,\beta_1)$ are unknown, sensitivity analysis helps evaluate how the estimates of parameters of interest change over a plausible range of values for $(\beta_0,\beta_1)$. 
Direct interpretation of $(\beta_0,\beta_1)$ may be challenging, while one can instead assess how the estimates change with different numbers of unpublished studies implied by $M$. 
Hereinafter, we refer to this sensitivity analysis method as the Copas-Shi method.

In addition,\cite{Copas2000} Copas proposed an selection equation related with the sizes of studies for dealing with PB based on the NN model.\cite{Copas1999} 
In the context of data summarized in contingency tables (such as Table \ref{tab:conf}), $\chi^2$ test is usually conducted to explore the effect of treatment on the outcomes, and the association between treatment and outcomes is measured by the phi coefficient, $\phi=\sqrt{\chi^2/n}$, which has an asymptotic variance of $Var(\phi)=1/n$. 
Thus, an alternative Gaussian variable was defined to model selective publication process, that is,
\begin{align}
Z_i = \gamma_0+\gamma_1\Ni^{1/2}+\delta_i,\label{sfzn}
\end{align}
where $\Ni=\Na+\Nb$ is the total samples size of study $i$, and $\delta_i$ is the residual term following $N(0,1)$.

With selective publication of studies, the log-likelihood conditional on the published studies was derived (see equation 7 in Copas\cite{Copas1999}); this method is referred to as the Copas-N method, hereinafter.
We presented a brief review of the Copas-N method in Section A of Supplementary Material.
However, the conditional log-likelihood of the Copas-N method is still limited to { the normal approximation, which requires continuity correction and is sensitive to it when the number of events includes zero.}
Moreover, in the absence of PB, the conditional log-likelihood does not exactly reduce to that of the NN model \eqref{rem}.

\section{Generalized linear mixed models for rare-event meta-analysis}\label{sec:glmm}

\subsection{Rare-event meta-analysis of odds ratios}
When the numbers of events in contingency table are close to or contain zero, the normal approximation using model \eqref{nn1} will be invalid for modeling the observed lnORs, leading to bias on the estimations of parameters of interest. 
Therefore, modeling the number of events using exact distributions is recommended.\cite{Stijnen2010}
For the estimation of the overall lnOR, Van Houwelingen et al.\cite{VanHouwelingen1993} and Stijhnen et al.\cite{Stijnen2010} proposed the HN model. 
Conditional on the total number of events ($y_i=\Ya+\Yb$) and the subjects of each group ($\Na$ and $\Nb$), the number of events in the treatment group is modeled by the Fisher's noncentral hypergeometric distribution:\cite{Fog2008a,Fog2008}
\begin{align}\label{hn1}
Y_{i1}\mid\thetai \sim \text{fnchypg}(\Na,\Nb,y_i,\exp(\thetai)),
\end{align}
where $\exp(\thetai)$ indicates the true study-specific ORs. 
Then, at the between-study level, the distribution of $\thetai$ is assumed to follow the normal distribution \eqref{nn2}.
By combing model \eqref{hn1} and \eqref{nn2}, the likelihood of the HN model is derived as 
\begin{align}\label{llkhn}
L(\theta,\tau)=\prodi \intR L_i(\thetai)\phidens d\thetai
\with
L_i(\thetai) =\dfrac{\binom{\Nb}{\Yb}\binom{\Na}{\Ya}\exp(\thetai\Yb)}{\sum_{k\in K} \binom{\Nb}{k}\binom{\Nb}{y_i-k}\exp(\thetai k)}.
\end{align}
where $K$ is the set of all possible values of $\Yb$ constrained by the marginal totals $\Na$ and $\Nb$ in Table \ref{tab:conf}, and $\theta$ is the overall lnOR of interest which can be estimated by the ML method with numerical integration.\cite{Stijnen2010}

When the number of subjects are large in the studies, estimation under the HN model, particularly the numerical integration, can be computationally demanding.\cite{Jansen2023} 
When the total occurrences of events are much smaller than the total subjects in the treatment and control groups, an approximation version of the HN model is possible to implement.
In this case, the probability \eqref{hn1} can be approximated by the following binomial distribution, which simplifies computation and model fitting:\cite{Stijnen2010,Jackson2018}
\begin{align}\label{bn1}
Y_{i1}\mid\thetai\sim\text{Bin}\left(y_i, \dfrac{\exp\{\log(\Nb/\Na)+\thetai\}}{1+\exp\{\log(\Nb/\Na)+\thetai\}}\right).
\end{align} 
The likelihood to estimate $\theta$ is constructed in the same way by replacing the within-study likelihood $L_i(\thetai)$ with that of the binomial distribution \eqref{bn1}, leading to the following conditional BN (CBN) model:\cite{Jansen2023}
\begin{align}\label{llkcbn}
L(\theta,\tau)=\prodi \intR L_i(\thetai)\phidens d\thetai
\with
L_i(\thetai) =\binom{y_i}{\Yb}\dfrac{\exp\{\log(\Nb/\Na)+\thetai\}^{\Yb}}{[1+\exp\{\log(\Nb/\Na)+\thetai\}]^{y_i}}.
\end{align}
The estimation performances of the HN and CBN models have been extensively studied, and the CBN model \eqref{llkcbn} is generally considered more computationally feasible than the HN model \eqref{llkhn}.\cite{Stijnen2010,Jansen2023,Jackson2018}

\subsection{Rare-event meta-analysis of proportions}
For meta-analyses involving single-group studies that estimates the overall log odds (equivalently, the logit-transformed proportion) of the event, the binomial distribution is employed to model the number of events in the single group (e.g., the treatment group in Table \ref{tab:conf}), that is,
\begin{align}\label{bn2}
Y_{i1}\mid\thetai\sim\text{Bin}\left(\Nb, \pi_i\right)
\with
\pi_i=\dfrac{\exp(\thetai)}{1+\exp(\thetai)}
\end{align}
where $\thetai=\logit(\pi_i)$, and $\pi_i$ represents the true occurrence probability of event in each study, 
Under the same framework, the likelihood is derived by combining the within-study likelihood of the binomial model \eqref{bn2} with the NN model \eqref{nn2}:
\begin{align}\label{llksbn}
L(\theta,\tau)=\prodi \intR L_i(\thetai)\phidens d\thetai
\with
L_i(\thetai) =\binom{\Nb}{\Yb}\dfrac{\exp(\thetai)^{\Yb}}{\{1+\exp(\thetai)\}^{\Nb}}.
\end{align}

To distinguish it from the CBN model \eqref{llkcbn} used for estimating the overall lnORs, we refer to this as the one-sample binomial-normal random-effects (1SBN) model.

\section{Sensitivity analysis for publication bias in generalized linear mixed models}\label{sec:sa}

In this section, we introduce the proposed sensitivity analysis methods that deal with PB in rare-events meta-analyses based on {the aforementioned HN and two BN models.}
Corresponding to Section \ref{sec:glmm}, we consider the situation of PB, where the $N$ published studies are regarded as biased sample from the population of related studies. 
To completely avoid continuity corrections for the zero entries and the estimation of SEs, it is natural to employ the latent Gaussian variable of the Copas-N method,\cite{Copas1999} as show in equation \eqref{sfzn}, to model selective publication process.
Given the observed sample sizes $\Ni$ in Table \ref{tab:conf}, we define the Gaussian variable $Z_i$ for study $i$ as
\begin{align}\label{sfnew}
Z_i = \alpha_0+\alpha_1{\Ni}^{1/2}+\delta_i 
\end{align}
where $\alpha_0$ and $\alpha_1$ are constants and $\delta_i$ is residual with distribution $N(0,1)$. 
Study $i$ is published if and only if $Z_i>0$, aligning with the interpretation of latent variable \eqref{sfzn}.
Based on the framework of the GLMM, it is difficult to re-express the observed outcomes using the standard normal residuals. 
Thus, we directly model the joint distribution of study-specific true effect sizes $\thetai$ with residuals $\delta_i$ in \eqref{sfnew}, expressed as follows:
\begin{align}\label{eq:corr}
\binom{\theta_i}{\delta_i}&\sim N\left( \binom{\theta}{0}, \smat{\tau^2}{\tau\rho}{\tau\rho}{1}\right),
\end{align}
where $\rho$ indicates the correlation between $\theta_i$ and $\delta_i$.
Then, the conditional distribution of $\delta_i$ given $\theta_i$ is derived as:
\begin{align*}
{\delta_i}\mid {\theta_i} \sim N\left(\rho\dfrac{1}{\tau}(\thetai-\theta), 1-\rho^2\right).
\end{align*}

On the other hand, the probability of study $i$ being published can be written by the probit model:
\begin{align}
P(Z_i>0\mid \Ni) = \Phi(\alpha_0+\alpha_1{\Ni}^{1/2}),\label{selectp1}
\end{align}
indicating that larger studies, or studies with significant results in $\chi^2$ test of the contingency table, are more likely to be published.
In the absence of PB, the likelihood conditional on the published studies with $Z_i>0$ is derived as 
\begin{align}\label{llk4}
L(\theta,\tau,\rho,\alpha_0,\alpha_1)=\prodi \int L_i(\theta_i\mid Z_i>0, \Ni)\phidens d\theta_i
\end{align}
with the conditional within-study likelihood being
\begin{align}\label{llk5}
L_i(\theta_i\mid Z_i>0,\Ni)& = \dfrac{ P(Z_i>0\mid \theta_i,\Ni) L_i(\theta_i)}{P(Z_i>0\mid\Ni)} \nonumber\\
&=\dfrac{P\left(\alpha_0+\alpha_1\sqrt{\Ni}+\delta_i>0\mid \theta_i,\Ni\right) L_i(\theta_i)}{P(Z_i>0\mid\Ni)} \nonumber\\
& = \dfrac{\Phi\left\{\dfrac{\alpha_0+\alpha_1\sqrt{\Ni}+\rho(\theta_i-\theta)/\tau}{\sqrt{1-\rho^2}}\right\}L_i(\theta_i)}{\Phi(\alpha_0+\alpha_1\sqrt{\Ni})},
\end{align}
where $L_i(\theta_i)$ indicates the exact within-study likelihoods introduced in Section \ref{sec:glmm}.
{When $\rho\neq0$, whether a study gets published is related with $\thetai$,} inducing PB when estimating the overall $\theta$.
When $\rho=0$, the conditional within-likelihood \eqref{llk5} reduced into $L_i(\thetai)$, implying that study outcomes are not influenced by any unpublished studies. 
In the Copas-N and Copas-Shi methods,\cite{Copas1999,Copas2000} the constant parameters are fixed as several values for sensitivity analysis.
Similarly, we also adopt the sensitivity analysis approach by fixing sensitivity parameters $(\alpha_0,\alpha_1)$ while estimating the other parameters $(\theta,\tau,\rho)$.
Since the sensitivity parameters $(\alpha_0,\alpha_1)$ are challenging in their interpretations, we considered their transformations in sensitivity analysis.
Let $\nmin=\min\{n_i\}$ and $\nmax=\max\{n_i\}$ be the minimum and maximum numbers of subjects among $N$ published studies;
$\Pmin$ and $\Pmax$ denote the probabilities of publishing a study with $\nmin$ and $\nmax$ subjects, respectively, (practically, $\Pmin \le \Pmax$).
With model \eqref{selectp1}, $\Pmin=\Phi(\alpha_0+\alpha_1\sqrt{\nmin})$ and $\Pmax = \Phi(\alpha_0+\alpha_1\sqrt{\nmax})$ hold. 
Given prespecified values of $(\Pmin, \Pmax)$, the constants $(\alpha_0,\alpha_1)$ are derived by 
\begin{align}\label{eq:alpha}
\alpha_1 = \dfrac{\Phi^{-1}(\Pmax) - \Phi^{-1}(\Pmin)}{\sqrt{\nmax}-\sqrt{\nmin}} \text{ and } 
\alpha_0 = \Phi^{-1}(\Pmax)-\alpha_1\sqrt{\nmax},
\end{align}
and then the other parameters $(\theta, \tau, \rho)$ in \eqref{llk4} are estimated by the ML method.
The number of unpublished studies is approximated in a similar way by\cite{Hattori2018} 
\begin{align*}
M=\sumi\dfrac{1-P(Z_i>0\mid\Ni)}{P(Z_i>0\mid \Ni)}.
\end{align*}

\section{Simulation studies}\label{sec:simu}
\subsection{Simulation designs}
Simulation studies were conducted to assess the performance of the proposed sensitivity analysis methods in adjusting PB in {the HN and two BN (CBN and 1SBN) models} under specified sensitivity parameters. 
We conducted separate simulations for meta-analysis of ORs and proportions. 

We considered meta-analyses with small and moderate numbers of population studies ($S=15,50$);
the population studies contained the published and unpublished studies. 
The overall true effect size, either the lnOR or the log odds, was set as $\theta=-2$.
The between studies variances were varied to reflect small to large heterogeneity, with values set as $\tau^2=0.1$, 0.3, or 0.7.
According to equation \eqref{eq:corr}, we generated the study-specific true effect size ($\thetai$) and residuals $\delta_i$.
For each study $i$, we considered different scenarios for the total subjects ($\Ni$), which were sampled from uniform distributions: $U[30,60]$, $U[50,200]$, and $U[500,700]$.
The scenarios with $U(500,700)$ might be impractical in real-world settings but was included to assess the approximation performance of the HN and CBN models under idealized conditions.
The total sample size $\Ni$ was then divided into treatment and control groups using different allocation ratios: $\Nb:\Na=1:1$ or $2:1$. 

In each meta-analysis, we simulated the data of each population study $i~(i=1,2,\dots,S)$ in the form of contingency table (Table \ref{tab:conf}) with small numbers of events.
For meta-analysis of ORs, we considered two data-generating processes to create the number of events in each group. 
The first generating process considered the HN model \eqref{llkhn} as the true data-generating mechanism.
Given a fixed total number of event ($y_i$) generated from $U[5,15]$ and the number of subjects in each group ($\Nb$ and $\Na$), the number of events in treatment groups ($\Ya$) were sampled from the hypergeometric distribution \eqref{hn1}; the remaining cells in the contingency table were derived accordingly.
{We referred to this as the HN model based data-generating process.}

The second generating process {reflected the design of prospective studies by separately generating the number of events in either group from binomial distributions; we referred to this process as the two-sample binomial normal (2SBN) model based data-generating process.}
Specifically, the probabilities of event for each study $i$ in the control group, denoted by $\p_{i0}$, were generated from a normal distribution $N(p_0,\tau^2/4)$ (following model 4 in Jackson et al.\cite{Jackson2018}), where three scenarios for $p_0=\{0.2,0.1,0.002\}$ were considered corresponding to different total sample size settings. 
By setting the true lnOR as $\theta=-2$, the event probability of each study in the treatment group ($\p_{i1}$) was derived accordingly.
Then, given the number of subjects in each group ($\Nb$ or $\Na$), the number of events in each group was independently generated from a binomial distribution: $y_{ij}\sim Bin(n_{ij}, p_{ij})$, where $j=\{0,1\}$, and $p_{ij}$ indicated the true probability of event in each group for study $i$. 

For meta-analysis of proportions in single group, it was difficult to generate rare events based on large number of subjects. 
Thus, we considered the scenarios of $\Ni$ sampled from $U[15,30]$ and $U[25,100]$ as the total number of subjects, the number of events were generated based on the BN mode \eqref{llksbn} given the true log odds being $\theta=-2$. {This is referred to as the 1SBN model based data-generating process.}
The summary of simulation scenarios for meta-analysis of ORs and proportions was presented in Table \ref{tab:setting1} and Table \ref{tab:setting2}, respectively.

{To simulate selective publication process}, we selected $N$ studies as the published ones from the $S$ population studies using selection function \eqref{sfnew}.
The detailed selection process is summarized as follows:
\begin{enumerate}
\item in each meta-analysis, given presepecified values on sensitivity parameters $(\Pmin,\Pmax)=(0.2,0.99)$, parameters $(\alpha_0,\alpha_1)$ were calculated according to equations \eqref{eq:alpha};
\item the random variable $Z_i$ was generated using equation \eqref{sfnew} for each studies given sample sizes ($\Ni$) and the generated $\delta_i$;
\item studies were selected as published ones if $Z_i>0$; otherwise, they were treated as unpublished.
\end{enumerate}

All simulation scenarios were repeated 1000 times, which means the performances of various methods were summarized based on 1000 meta-analyses.
We evaluated the performance of the NN, HN, and BN (CBN and 1SBN) models using both the full population studies (published and unpublished studies) and the subset of published studies only.
{In the NN model, the continuity correction was conducted by adding 0.5 to each cell of the contingency table for the studies with zero entries.}
The estimates based on the full population and the published studies were denoted by subscripts $P$ and $O$, respectively (e.g., HN$_P$, HN$_O$), and the differences between the two types of estimates reflected the magnitude of PB under each model.
{To assess the ability of adjusting PB, we compared the performance of the propose methods with the Copas-N\cite{Copas1999} and Copas-Shi\cite{Copas2000} methods which adopted the Copas-Heckman-type selection functions and rely on the NN model. }
In sensitivity analyses, sensitivity parameters were set to their true values for estimating the other parameters of interest. 
While this strategy is not feasible in real-world applications, it was used exclusively in simulation studies.
To differentiate results derived from the various models, we denoted the proposed method with the superscript Prop (e.g., HN$^\text{Prop}$)

The simulation studies were implemented by R with further details provided {in Section B.1 of Supplementary Material.}
The reproducible R code is available on GitHub.

\subsection{Result 1: meta-analysis of odds ratios under the HN model based data-generating process}

Under all scenarios, a large proportion of studies included rare events (defined as fewer than 3 events in either group).
The occurrences of rare events in the population and published studies were summarized and presented respectively {in the HN$_P$ and HN$_O$ columns of Table B1 of Supplementary Material.}
The estimates of the overall lnOR (i.e., $\theta$) were summarized in Table \ref{tab:set1}.
When population data were generated by the HN model, the estimates of the HN model using population studies (HN$_P$) yielded minimal bias, as expected.
In contrast, the estimates of NN model (NN$_P$) showed substantial bias across all rare-event meta-analysis scenarios.
When the numbers of studies and patients in studies were sufficiently large, estimates by the CBN model (CBN$_P$) tended to have reduced bias and closely approximated the HN model results.
When estimations were based only on the published studies, substantial PB was observed across all the three models, reflected in the biased estimates NN$_O$, HN$_O$, and CBN$_O$.
The magnitude of PB was especially increased when the between-study heterogeneity parameter $\tau$ was moderately large.

{In the sensitivity analyses, the Copas-Shi method generally resulted in large bias.}
In contrast, the proposed HN model based method (HN$^\text{prop}$) achieved small bias, especially when $\tau$ is neither very large nor small, indicating that the proposed method successfully adjusted PB.
The proposed CBN model based method (CBN$^\text{prop}$) derived larger bias than HN$^\text{prop}$; however, when subjects were large and $\tau$ was small, the bias in the CBN$^\text{prop}$ became decreased.
Notably, the Copas-N method also derived small bias; however, when $\tau$ was small and subjects in the two groups were unbalanced, the bias was non-negligible. 
The proposed HN model based method appeared robust against group imbalance, whereas the other methods showed increased bias when imbalance was present.
{The proposed methods achieved coverage probabilities closer to the nominal level than the competing methods,} although some coverage probabilities did not reach nominal levels, possibly due to imperfect specification of $(\Pmin, \Pmax)$ in the sensitivity analyses.

The CBN model was introduced to reduce computational complexity and mitigate non-convergence issues.
Therefore, we investigated the proportions of estimations that successfully converged (out of 1000 runs), summarized as convergence rates {in Table B3 of the Supplementary Material}.
Results showed that the NN model had near-perfect convergence rates, whereas the HN and CBN models, especially in the proposed methods, showed comparatively lower convergence rates; however, when the number of studies were not too small, satisfactory convergence rates could still be obtained.
On the other hand, convergence rates were comparable between the HN and CBN models across the simulation settings.
The estimation of $\tau$ were additionally investigated and summarized {in Table B4 of Supplementary Material}. 
The proposed methods demonstrated less bias in estimating $\tau$, while the Copas-N method tended to overestimate and the Copas-Shi method tended to underestimate $\tau$.

\subsection{Result 2: meta-analysis of odds ratios under the 2SBN model based data-generating process}

Under the generating process of 2SBN model, large proportions of rare events were derived, the the proportions decreased when the number of total subjects were large. However, the rare events still accounted for relatively large proportions in the population and published samples, shown as the 2SBN$_P$ and 2SBN$_O$ in {Table B1 of Supplementary Material}.
The average estimates of the overall lnOR, $\theta$, in Table \ref{tab:set2} showed that the HN model (both HN$_P$ and HN$^\text{prop}$) generally produced less bias than other methods; with bias tending to decrease as $\tau$ became smaller.
Estimates from the CBN$_P$ and CBN$^\text{prop}$ models were larger than those from the HN models but tended to decrease as the number of patients increased.
When $\tau$ was small, the bias in CBN$_P$ and CBN$^\text{prop}$ could be decreased. 
Under this data-generating process, both the Copas-N and Copas-Shi methods failed to adequately adjust for PB, with the Copas-N method exhibiting greater bias.
Compared to the Copas-N and Copas-Shi methods, {the proposed methods, especially the HN model based method, achieved coverage probabilities closer to the nominal level} for estimating the lnOR.

The convergence rates showed that the HN and BN models had comparable performance in estimation convergence ({Table B5 in Supplementary Material}).
The estimation results of $\tau^2$ indicated that the HN and BN models and the proposed methods gained less bias when the number of subjects and studies were large ({Table B6 in Supplementary Material}). In contrast, overestimation and underestimation of $\tau^2$ were found in the Copas-N and Copas-Shi methods, respectively.

\subsection{Result 3: meta-analysis of proportions under the 1SBN model based data-generating process}

The scenarios of generating single-group data were presented in Table \ref{tab:setting2}. When the number of subjects were large, it is difficult to generate large proportions of rare events in the simulated studies; however, the amount of rare events still accounted for approximately 20\% and more, as shown in {Table B2 of Supplementary Material}. 
The average estimates of the log odds (i.e., $\theta$) were summarized in Table \ref{tab:set3}.
The results indicated that the 1SBN model yielded less bias than the NN model based on population studies; however, when the number of subjects was sufficiently large and rare events were not large, both models showed comparable performance with minimal estimation bias.
Estimates based on published studies showed that PB increased with the increasing magnitude of $\tau$.
In the sensitivity analysis, the proposed 1SBN model (1SBN$^\text{prop}$) based method achieved less bias and {coverage probability closer to the nominal level overall,} indicating successful adjustment for PB.
On the other hand, the Copas-Shi method exhibited large bias and failed to adequately adjust for PB, while the Copas-N method tended to perform better with less bias when the sizes of studies were large.

All methods generally achieved good convergence performance ({Table B7 in Supplementary Material});
when the number of studies and the number of total subjects were small, the 1SBN model and the proposed method had some deficiencies in convergence rates. 
The summary of estimation of $\tau^2$ ({Table B8 in Supplementary Material}) showed similar patterns: the proposed methods had less bias, while the Copas-N method tended to overestimate and the Copas-Shi method tended to underestimate $\tau^2$.

\section{Application}\label{sec:app}

\subsection{Example 1: rare-event meta-analysis of odds ratios}\label{sec:eg1}
We revisit the meta-analysis of lnORs in Stijhnen et al.\cite{Stijnen2010} 
This is the meta-analysis of lnORs of catheter-related bloodstream infection (CRBSI) between anti-infective-treated (AIT) central venous catheters and the standard catheter.\cite{Niel-Weise2007}
The data of this meta-analysis were presented in {Table C9 of Supplementary Material,} where five studies had zero event in the AIT catheter and one study had zero entries in both standard and AIT catheters.
Stijnen et al.,\cite{Stijnen2010} compared meta-analytical results by the NN model with those by the HN and CBN models without considering selective publication of studies.  
We reproduced the estimations of parameters in various models by maximizing likelihoods \eqref{llk1}, \eqref{llkhn} and \eqref{llkcbn}. 
The SEs of the parameters were estimated by the inverse of the empirical Fisher information matrix following the ML theory.
The HN and CBN models estimated the overall lnOR with 95\% CI to be $-1.353~(-2.041, -0.665)$ and $-1.303~(-1.966, -0.639)$, respectively, which were consistent with the estimations reported by Stijnen et al.\cite{Stijnen2010}
The approximation in the results was attributed to the nature of the data, where the total number of CRBSI was relatively small to the total number of subjects.
The NN model overestimated the overall lnOR to be $-0.955~(-1.415, -0.495)$, which turned to be biased towards zero.\cite{Stijnen2010} 
The results showed that there were certain discrepancies between the NN model and the HN or CBN model, indicating that normal approximation in the NN model was questionable. 
These estimates derived by the ML method were identical with those generated by R package \texttt{metafor}. 
The estimates of the NN model could be reproduced by R function \texttt{rma} with \texttt{method=``ML''}, while the estimates of the HN and CBN models were identical with \texttt{rma.glmm} by setting \texttt{model=``CM.EL''} and \texttt{model=``CM.AL''}, respectively.

{To evaluate PB in this meta-analysis, we firstly employed the trim-and-fill method to adjust for the potential impact of PB. At the same time, Egger's regression and Begg's rank tests\cite{Egger1997} were performed to assess funnel plot asymmetry, which suggested the presence of PB.}
{These methods were developed based on the NN model. When there were zero entries, we conducted two approaches of continuity correction for the data: (a) adding 0.5 to all the cells of contingency tables for studies with zero entries only; (b) adding 0.5 for all the cells of all the studies.}   
The results of trim-and-fill methods and the regression and rank tests corresponding to the two continuity correction methods were shown in Figure \ref{fig:eg1}A and B, respectively.
In Figure \ref{fig:eg1}A, seven studies were filled, while in Figure \ref{fig:eg1}B, six studies were filled. 
{After accounting for the imputed studies and applying continuity correction approaches (a) and (b), the estimates and 95\% CIs of the lnOR were adjusted from $-0.955~(-1.415, -0.495)$ to $-0.513~(-1.010, -0.017)$ and from $-0.861~(-1.283, -0.440)$ to $-0.506~(-0.980, -0.031)$, respectively.
On the other hand, neither Egger's regression and Begg's rank tests indicated significant asymmetry of the funnel plots.}

The adjustments of PB were based on the symmetry of funnel plot and required continuity correction, which might be subjective and biased for rare-event meta-analyses. Thus, we evaluated the potential PB on the lnOR by the proposed methods, which was free from continuity correction, and compared the results with other existing methods, including the Copas-Heckman-type sensitivity analysis methods\cite{Copas1999,Copas2000} and the $t$-statistic selection function based sensitivity analysis methods\cite{Hu2024} (hereinafter, the $t$-statistic method) under both approaches of continuity correction.
The $t$-statistic methods\cite{Hu2024} were also developed based on the HN and BN models; here, we only considered the exact HN model based one.
The parameters $(\theta, \tau, \rho)$ were then estimated by the ML method given presepcified values of sensitivity parameters.
For the proposed method and the Copas-N method,\cite{Copas1999} we set $(\Pmin, \Pmax) = \{(0.99, 0.999), (0.9, 0.999), (0.8, 0.999), \dots, (0.1, 0.999)\}$, where $\Pmin$ and $\Pmax$ were probabilities of publishing studies with the largest and smallest numbers of subjects, respectively. 
For the Copas-Shi method,\cite{Copas2000} the sensitivity parameters $\Pmin$ and $\Pmax$ indicated the probability of publishing studies with minimum and maximum SEs, respectively; then $(\Pmin, \Pmax)$ were set as $\{(0.999, 0.99), (0.999, 0.9), (0.999, 0.8), \dots, (0.999, 0.1)\}$.
For the $t$-statistic method,\cite{Hu2024} the sensitivity parameter $p$ was defined as the marginal probability of publishing studies from the population; thus, $p=1,0.9,0.8,\dots, 0.1$ were set.
When the probabilities in the sensitivity parameters tended to 1, the approximated numbers of unpublished studies were close to 0, indicating the situation of no PB. 

The impact of PB was assessed via the changes of the estimated lnOR under various sensitivity analyses, and the results were presented in Figure \ref{fig:eg1}C-D with the detailed estimations summarized {in Table C10-13 of Supplementary Material}.
Given the fixed value of $\Pmax = 0.999$, $\Pmin$ decreasing from 0.99 to 0.1 (shown by $x$-axis in Figure \ref{fig:eg1}C-F) assumed that the number of unpublished studies increased, as illustrated on the plots.
By the proposed HN and CBN model based methods, the estimated lnOR increased subtlety with increasing number of unpublished studies, indicating that PB had little influence on the estimation of the overall lnOR (Figure \ref{fig:eg1}C). This conclusion was consistent with the results of funnel plots.  
On the other hand, the Copas-N method overestimated the overall lnOR at each $\Pmin$ (Figure \ref{fig:eg1}D), and the Copas-Shi method based on the NN model moreover induced an overestimation on the magnitude of PB (Figure \ref{fig:eg1}E). Both methods were influenced by the different approaches of continuity correction. 
The $t$-statistic method assessed potential PB using selection function related with the $t$-statistic (equivalently, $p$-value) of the lnORs; it also gave a reasonable evaluation of PB and indicated that the estimated overall lnOR tended to be insignificant when the number of missing studies were greater than four; however, the computational time was about 6 times of the propose methods. 
We summarized the computational time of the proposed methods and the $t$-statistic method in {in Table C28 of Supplementary Material}.

According to the results of the proposed methods, in summary, the estimations of the lnOR remained significant in the presence of unpublished studies, indicating that the meta-analytical lnOR estimated by the original HN and CBN models were robust against potential PB in this example.


\subsection{Example 2: rare-event meta-analysis of odds ratios}\label{sec:eg2}

The second example included 16 trials examining the effectiveness of intravenous magnesium in the prevention of death following acute myocardial infarction, where eight studies had less-than-three events in the magnesium group and four studies in the control group. 
The data of this meta-analysis were presented in {Table C14 of Supplementary Material.}

In the absence of PB, the lnOR was estimated to be $-0.844 (-1.298, -0.390)$ by the HN model, $-0.752 (-1.177, -0.327)$ by the CBN model, and $-0.746 (-1.194, -0.299)$ by the NN model, respectively. 
To evaluate the impact of PB, various methods were compared with the results presented in Figure \ref{sec:eg2}. 
Under two approaches of continuity correction, the funnel plots with Egger's test of asymmetry indicated the existence of PB; the trim-and-fill methods estimated seven unpublished studies and adjusted the overall lnOR from $-0.746$ to $-0.375 (-0.791, 0.042)$ and from $-0.691$ to $-0.348 (-0.727, 0.032)$, respectively (Figure \ref{sec:eg2}A and B). 
However, the NN model and the trim-and-fill method had risks of overestimating the lnOR.

On the other hand, the proposed methods implied that the overall lnOR might become insignificant when there were more than 19 studies unpublished (Figure \ref{sec:eg2}C). The estimates by the Copas-Heckman-type methods and the $t$-statistic method were slightly influenced by the different approaches of continuity correction.
Among these results, the Copas-N method (Figure \ref{sec:eg2}D) underestimated of the overall lnOR in the absence of PB (equivalently, $\Pmin\approx 1$), while the Copas-Shi method (Figure \ref{sec:eg2}E) tended to overestimate the lnOR and yield unstable estimations when the number of unpublished studies were large. 
In this example, the $t$-statistic method derived unstable estimations and very long computation duration; the analysis of this method implied a decreasing estimation of the lnOR when the number of missing studies increased, as shown in Figure \ref{sec:eg2}F.
The detailed estimates by these methods were presented in {in Table C15-18 of Supplementary Material}.

\subsection{Example 3: rare-event meta-analysis of proportions}\label{sec:eg3}

{In this case, we only adopted the data in the AIT catheter group, following the analysis by Stijnen et al;}\cite{Stijnen2010}the parameter of interest is the overall log odds of CRBSI.
The number of events ranged from 0 to 6 while the number of subjects ranged from 44 to 345.
Without considering PB, the log odds of CRBSI and its 95\% CI were estimated to be $-4.812~(-5.508,  -4.116)$ and $-4.238~(-4.793,  -3.682)$ by maximizing the likelihoods of the 1SBN model \eqref{llkcbn} and the NN model \eqref{llk1}, respectively.

To evaluate the impact of PB on the log odds of CRBSI, we employed the proposed method and compared the results with those by the Copas-Heckman-type methods and the $t$-statistic method.
We adopted the same scenarios on the sensitivity parameters, $(\Pmin, \Pmax)$ and $p$, as described in Section \ref{sec:eg1}, to re-estimate the parameters of interest. 
The sensitivity analysis results by various methods were presented in Figure \ref{fig:eg3}.

{The funnel plot with Egger's test revealed significant asymmetry of funnel plot, implying the existence of PB. Furthermore, the trim-and-fill method indicated that eight and seven studies were potentially unpublished under two approaches of continuity correction, respectively.} The log odds were estimated to be increased with accounting for the unpublished studies.
By the proposed BN method, as the increasing number of unpublished studies, the estimated overall log odds were decreased (Figure \ref{fig:eg3}C), {indicating that ignoring selective publication of studies might induce overestimation of the meta-analysis results.}
The Copas-N method also estimated a decreasing log odds with increasing number of unpublished studies, where the estimates decreased abruptly when $\Pmin$ was less than 0.6 (Figure \ref{fig:eg3}D).
In contrast, the Copas-Shi method and the $t$-statistic method yielded increasing estimations of the log odds (Figure \ref{fig:eg3}E and F) when assuming that PB was related with the SEs and the $p$-value of the log odds.
Among the two methods, the Copas-Shi method had apparent overestimations (Figure \ref{fig:eg3}E).

Although the signs of PB were estimated differently, the impact of PB analyzed by all the sensitivity analyses were not significant, the detailed estimations were summarized {in Table C19-22 of Supplementary Material}. 
The results indicated that the overall log odds of CRBSI was subtlety influenced by PB. Thus, a robust conclusion from the meta-analysis could be drawn.

\subsection{Example 4: rare-event meta-analysis of proportions}\label{sec:eg4}

This meta-analysis originally aimed to estimate the overall success rate of hyperdynamic therapy for treating cerebral vasospasm and included 14 studies on the effectiveness of hyperdynamic therapy.\cite{Zhou1999}
To provide a more intuitive explanation, we define the event of interest to be the not improved, and then two studies had 0 events and six studies had less-than-three events.
The data of this example were presented {in Table C23 of Supplementary Material}. 
In this case, the parameter of interest was the log odds of overall failure rate of hyperdynamic therapy.
Without considering PB, the log odds with 95\% CI was estimated to be $-1.377~(-1.942, -0.811)$ by the 1SBN model and $-1.118~(-1.598, -0.637)$ by the NN model.

The same methods were implemented to assess the impact of PB on the results, as shown in Figure \ref{sec:eg4}.
Under two approaches of continuity correction, Egger's regression test did not detected significant existence of PB in the funnel plots, and the trim-and-fill methods indicated four and three studies, respectively. With the filled methods, the larger estimations on the log odds were adjusted (Figure \ref{sec:eg4}A and B).
The sensitivity analysis methods presented the adjusted estimates of the log odds given various numbers of unpublished studies. 
The proposed method and {the Copas-Shi and Copas-N methods} derived increasing estimates of the log odds while the $t$-statistic method showed the decreasing estimate, when the number of unpublished studies increased (Figure \ref{sec:eg4}C-F).
However, the Copas-N method seemed to underestimate the overall log odds while the Copas-Shi method overestimated it (Figure \ref{sec:eg4}D and E). 
The $t$-statistic method required large amount of time and derived some unstable estimations (Figure \ref{sec:eg4}F).
Except for the proposed method, the other three methods were affected to some extent by different approaches continuity correction.
The detailed estimations were summarized {in Table C24-26 of Supplementary Material}. 
The proposed sensitivity analysis method provided increased confidence that the overall log odds of the failure rate estimated by the 1SBN model is robust to PB in this example.

\section{Discussion}\label{sec:diss}
\label{sec5}

Increasing the estimation accuracy for rare-event meta-analysis has been extensively studied over the years.
Various models employing the exact likelihoods of events have been proposed to increase the estimation accuracy, including the beta-binomial models\cite{Kuss2015,Mathes2018,Xu2022} and Poisson random-effects model.\cite{Cai2010}
Among these, the HN and BN models appear to have relatively broad applications, since they have been implemented in widely used R packages for meta-analyses,\cite{Jackson2018} such as \texttt{metafor}\cite{metafor} and \texttt{meta}.\cite{meta}
In systematic review and meta-analyses, PB has been recognized as an inevitable issue.
According to the Preferred Reporting Items for Systematic reviews and Meta-Analyses (PRISMA) guideline,\cite{Page2021} it is recommended to conduct sensitivity analyses to assess the robustness of the synthesized results (item 13f), and to present assessments of bias due to missing results (item 21).
To deal with PB, many advanced selection-model-based methods have been proposed.\cite{Copas1999,Copas2000,Copas2001,Ning2017,Huang2021,Li2022,Huang2023} However, implementing these methods for rare-event meta-analyses can be challenging and problematic since they are developed based on the NN model.

The proposed methods serve as useful tools for addressing PB and assessing the robustness of results synthesized by the HN and BN models, overcoming the lack of methods for addressing PB in rare-event meta-analyses.
The proposed methods employed the selection function on the sample size of a study, which was firstly proposed by Copas.\cite{Copas1999}
Recently, this selection function was also employed to deal with PB in meta-analysis of diagnostic studies.\cite{Hattori2018}
In Copas-Heckman-type sensitivity analyses, the outcomes were linked with the Gaussian latent variable modeling selective publication process, while in the proposed method the study-specific true effects sizes were directly linked for addressing PB.
Simulation studies have shown that the proposed methods, especially the HN model based sensitivity analysis method, obtained satisfactory performance in adjusting PB in the meta-analytical estimates. 
A recently proposed method for dealing with PB also proposed based on the framework of GLMM, where the selection function on $t$-statistic of the lnOR was considered.\cite{Hu2024} 
However, the continuity correction was still required for calculating the $t$-statistics, and the Heckman-type selection function utilized in this paper cannot be handled.
The computation times summarized for the examples further demonstrated that this method was considerably more time-consuming, especially with larger number of subjects and studies.
Our proposal totally avoid the continuity correction and is much simper in the estimation.
{The proposed method is not intended to replace the $t$-statistic selection function based sensitivity analysis method.}\cite{Hu2024} 
In practice, since the explicit selecting mechanism behind PB is unknown, it is recommended to conduct sensitivity analyses with different selection functions to evaluate PB and draw robust conclusions.

The proposed methods are advantageous in their applicability, as they can be applied to any rare-event meta-analyses that can be synthesized using the HN or BN models, and can be easily implemented by free statistical software, such as R.
The proposed methods handle PB in rare-event meta-analyses of different outcomes, and can be extended into similar models, such as meta-analysis of incidence rate ratio using the BN model.\cite{Stijnen2010}
We regard the proposed methods as supplemental analyses aiming at evaluating PB in rare-event meta-analysis.
For rare-event meta-analysis of odds ratios, we suggest using the HN model for meta-analysis and the proposed HN model based sensitivity analysis for PB evaluation. 
Although the HN based sensitivity analysis method involves some computational complexity, the computation time is typically within a few minutes or less when the number of studies is moderate.
The proposed BN based method is recommended only when the number of studies and subjects is sufficiently large, as it can help reduce computation time in such cases.

{Several studies have indicated that the beta-binomial model exhibits better performance in rare-event meta-analyses.\cite{Kuss2015,Mathes2018,Xu2022} 
However, extending the proposed framework into this model is not straightforward and will be considered in future work.
In some cases, rare events coincide with meta-analyses that comprise only a few studies.
Our simulation studies demonstrated that the propose methods worked in meta-analysis with about ten published studies with less bias in the HN model based method. 
If the number of studies is too small, the proposed methods may suffer from convergence issues and the increased estimation bias.
Despite these limitations, our proposal is still valuable and provides a straightforward and practical approach to assessing PB in rare-event meta-analyses using the GLMMs.


\bmsection*{Acknowledgments}

{This research was partly supported by Grant-in-Aid for Challenging Exploratory Research (16K12403) and for Scientific Research (16H06299, 18H03208) from the Ministry of Education, Science, Sports and Technology of Japan.}
This work was (partly) achieved through the use of {SQUID and/or ONION} at D3 Center, The University of Osaka.

\bmsection*{Data availability statement}
R codes used in this paper together with a sample application data set are available at
\url{https://github.com/meta2020/remetasa-r}

\bmsection*{Supporting information}
Additional supporting information can be found online in the Supporting Information section at the end of this article.

\bibliography{refs}%

\clearpage


\begin{figure}[!htbp]
\begin{center}
\centerline{\includegraphics[width=\textwidth]{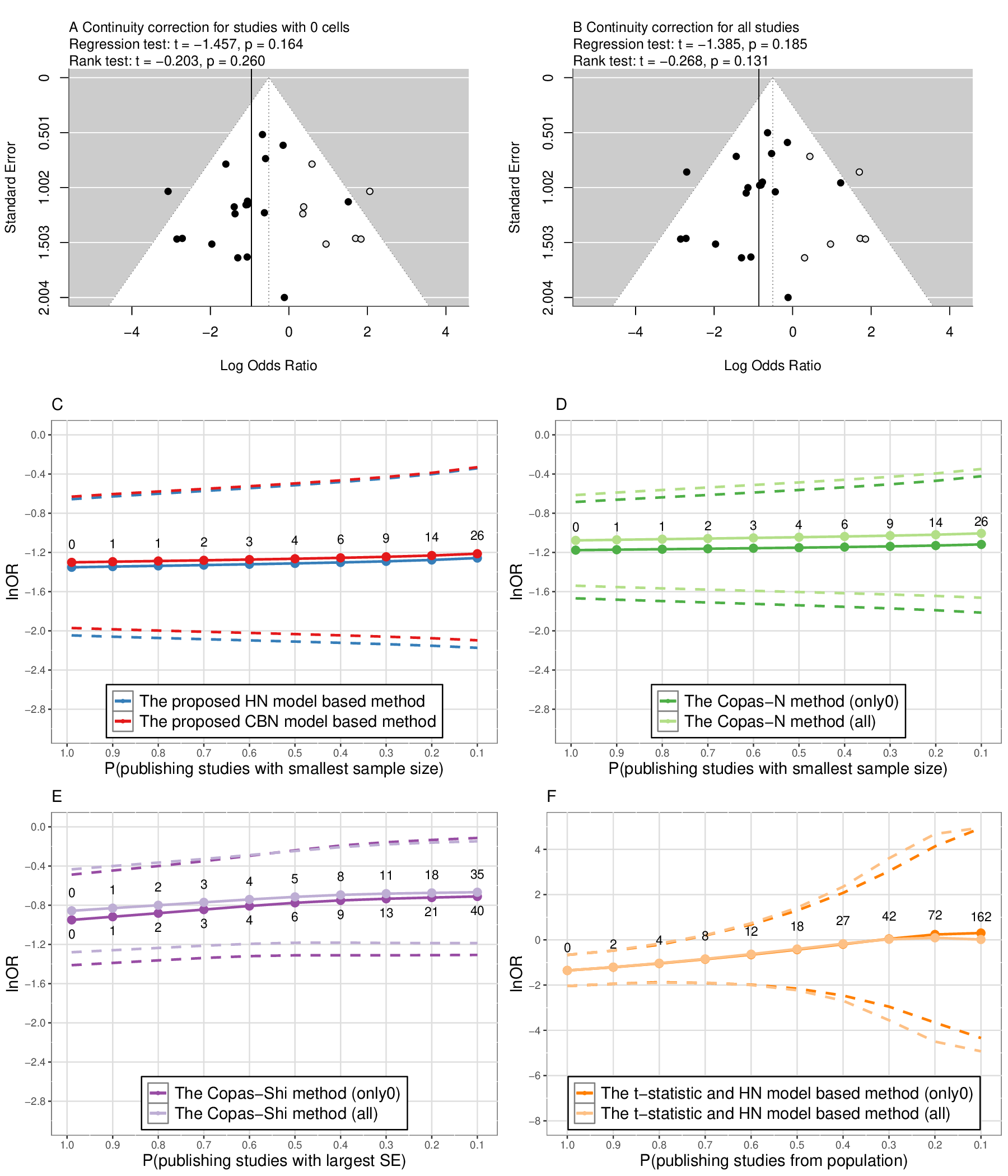}}
\end{center}
\caption{Comparison of sensitivity analysis methods for PB in Example 1. 
The values indicate numbers of missing studies.
Methods in panels (B), (C), and (D) need continuity correction; only0 indicates continuity correction for studies with 0 cells only, while all indicates continuity correction for all studies.}\label{fig:eg1}
\end{figure}


\begin{figure}[!htbp]
\begin{center}
\centerline{\includegraphics[width=\textwidth]{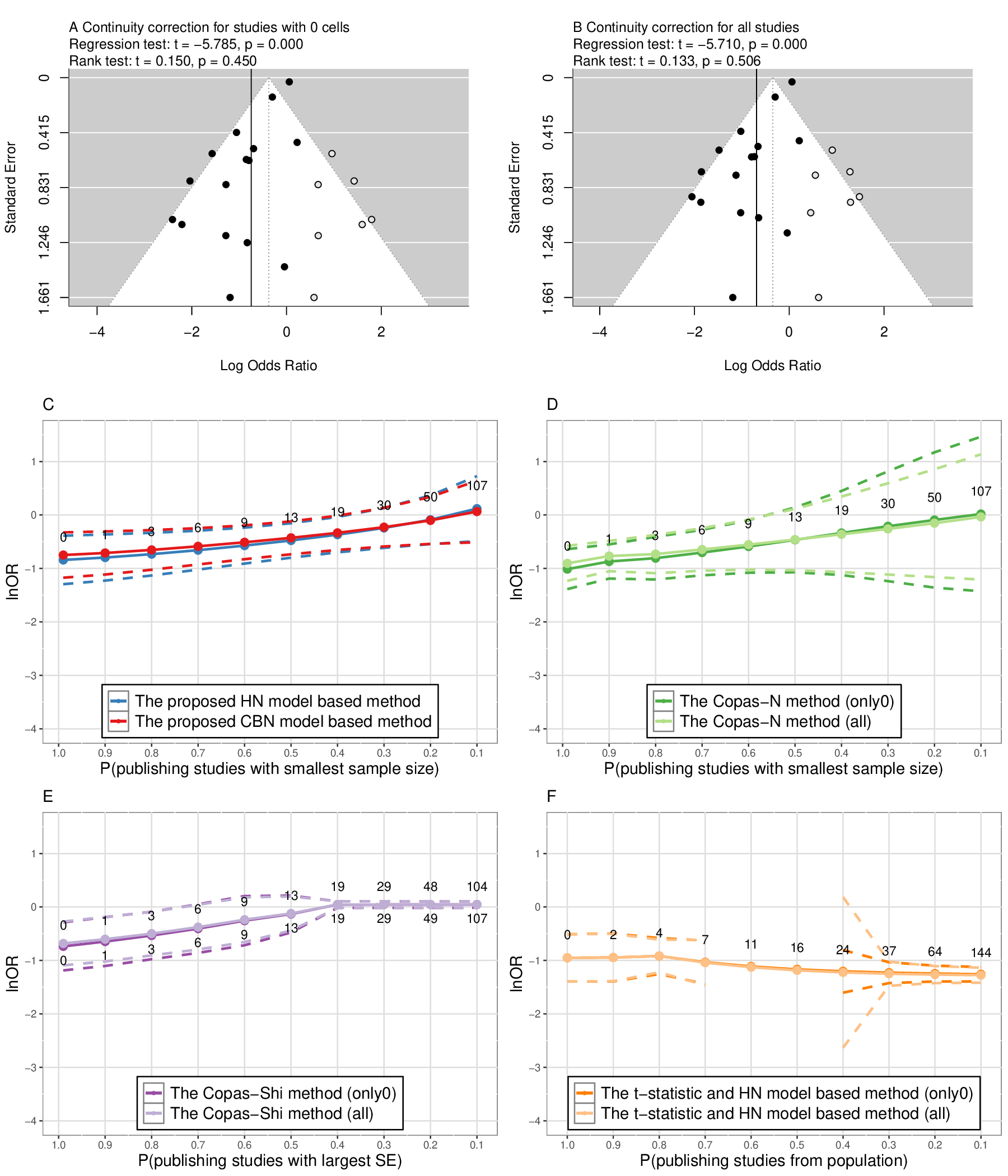}}
\end{center}
\caption{Comparison of sensitivity analysis methods for PB in Example 2. 
The values indicate numbers of missing studies.
Methods in panels (B), (C), and (D) need continuity correction; only0 indicates continuity correction for studies with 0 cells only, while all indicates continuity correction for all studies.}\label{fig:eg2}
\end{figure}


\begin{figure}[!htbp]
\begin{center}
\centerline{\includegraphics[width=\textwidth]{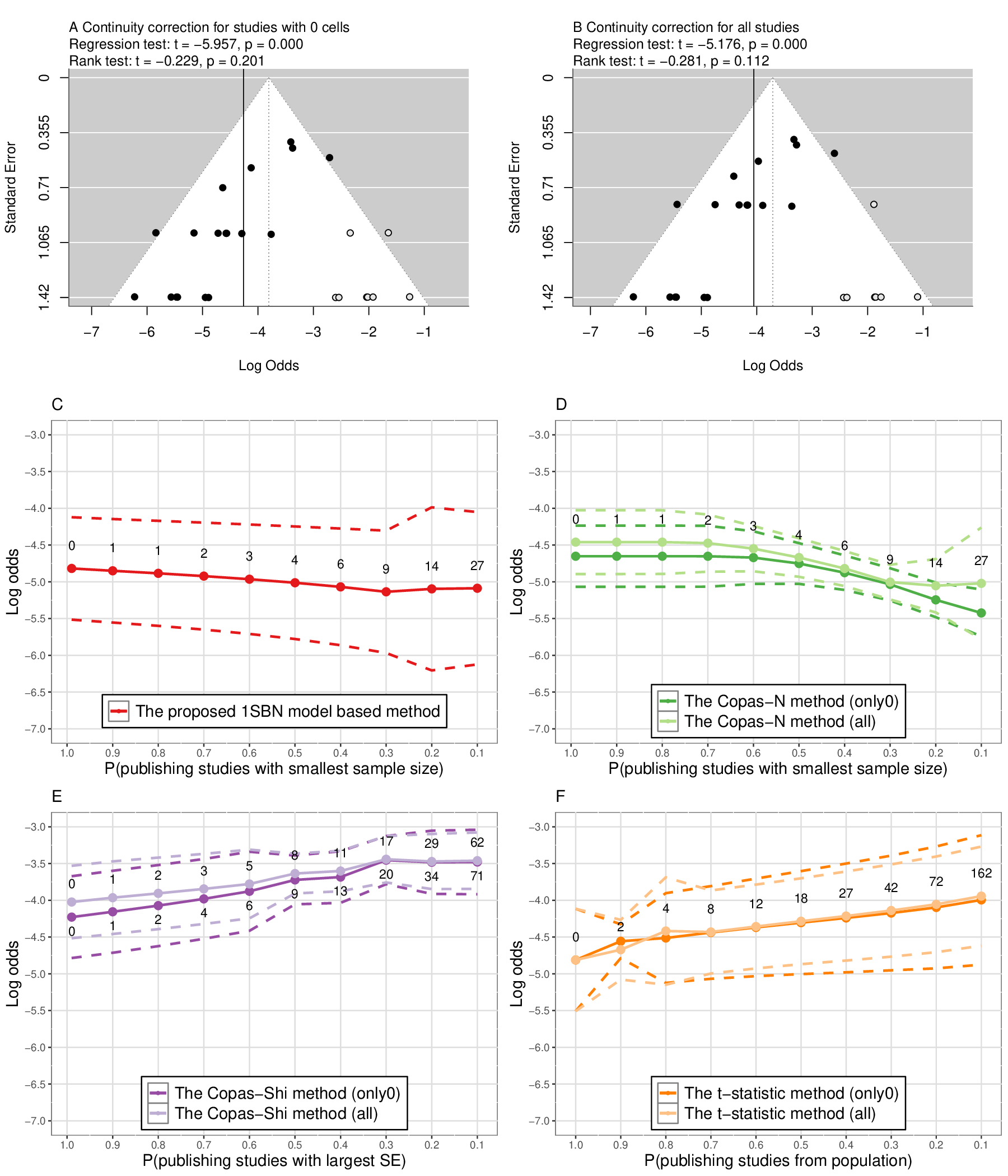}}
\end{center}
\caption{Comparison of sensitivity analysis methods for PB in Example 3. 
The values indicate numbers of missing studies.
Methods in panels (B), (C), and (D) need continuity correction; only0 indicates continuity correction for studies with 0 cells only, while all indicates continuity correction for all studies.}\label{fig:eg3}
\end{figure}

\begin{figure}[!htbp]
\begin{center}
\centerline{\includegraphics[width=\textwidth]{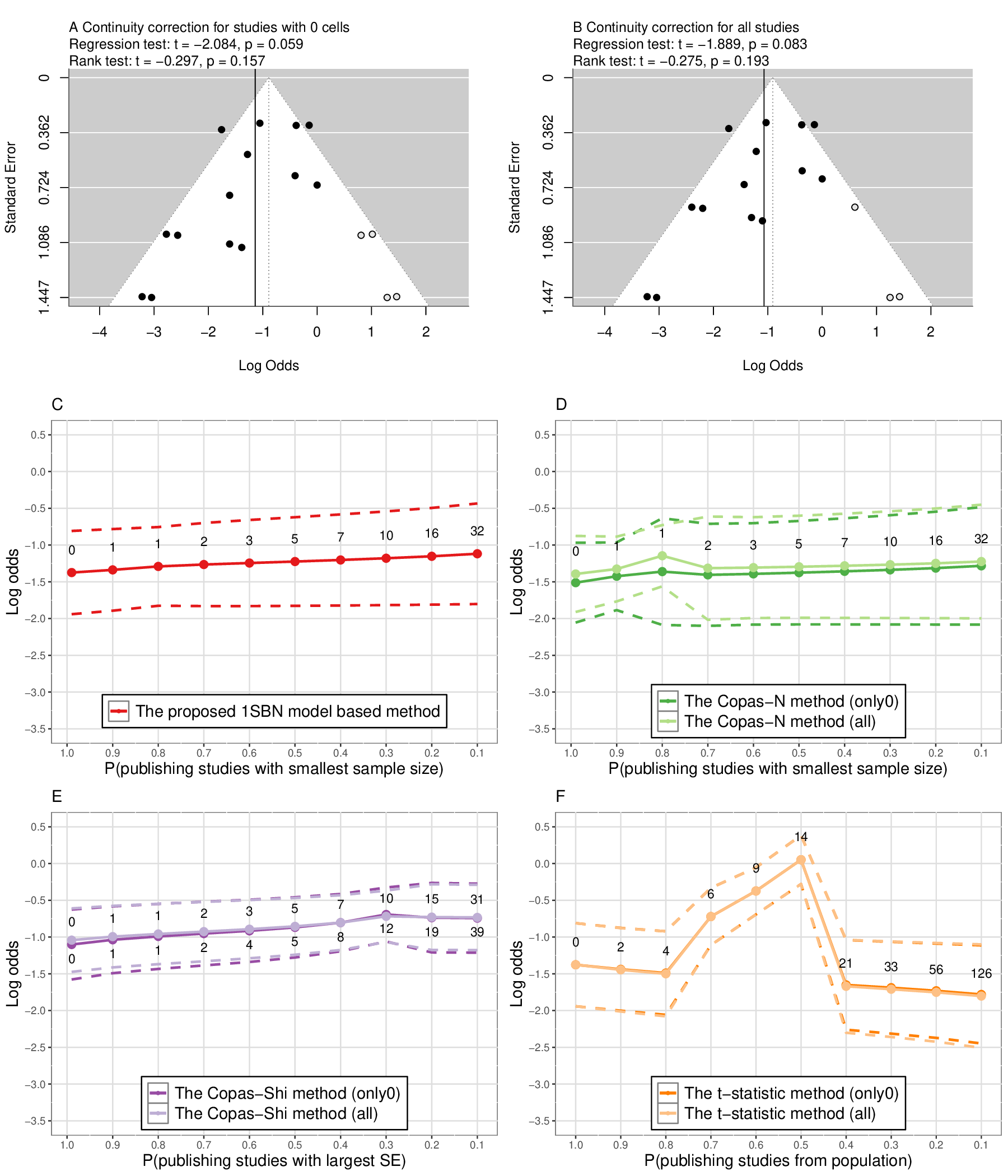}}
\end{center}
\caption{Comparison of sensitivity analysis methods for PB in Example 4. 
The values indicate numbers of missing studies.
Methods in panels (B), (C), and (D) need continuity correction; only0 indicates continuity correction for studies with 0 cells only, while all indicates continuity correction for all studies.}\label{fig:eg3}
\end{figure}

\clearpage

\begin{table}[!htbp]
\caption{Example of the contingency table for data of study $i$}
\centering
\label{tab:conf}
\begin{tabular}{@{}rrrr@{}}
\toprule
          & Event & Non-event & Total \\ \midrule
Treatment & $\Yb$    & $\Nb-\Yb$    & $\Nb$    \\
Control   & $\Ya$    & $\Na-\Ya$    & $\Na$    \\ \midrule
Total     & $\Yi$    & $\Ni-\Yi$    & $\Ni$    \\ \bottomrule
\end{tabular}%
\end{table}

\begin{table}[!htbp]

\caption{\label{tab:setting1}Scenarios for simulating meta-analysis of odds ratios}
\centering
\begin{threeparttable}
\begin{tabular}[t]{rrrrrrrr}
\toprule
True lnOR ($\theta$) & $\rho$ & Total event ($y_i$) & $(P_{max}, P_{min})$ & Total subjects ($n_i$) & $p_0$ & T:C & $\tau^2$\\
\midrule
-2 & 0.8 & {}[5, 15] & (0.99, 0.20) & {}[30, 60] & 0.2 & 1:1 & 0.1\\
 &  &  &  &  &  &  & 0.3\\
 &  &  &  &  &  &  & 0.7\\
\addlinespace
 &  &  &  &  &  & 2:1 & 0.1\\
 &  &  &  &  &  &  & 0.3\\
 &  &  &  &  &  &  & 0.7\\
\addlinespace
 &  &  &  & {}[50, 200] & 0.1 & 1:1 & 0.1\\
 &  &  &  &  &  &  & 0.3\\
 &  &  &  &  &  &  & 0.7\\
\addlinespace
 &  &  &  &  &  & 2:1 & 0.1\\
 &  &  &  &  &  &  & 0.3\\
 &  &  &  &  &  &  & 0.7\\
\addlinespace
 &  &  &  & {}[500, 700] & 0.002 & 1:1 & 0.1\\
 &  &  &  &  &  &  & 0.3\\
 &  &  &  &  &  &  & 0.7\\
\addlinespace
 &  &  &  &  &  & 2:1 & 0.1\\
 &  &  &  &  &  &  & 0.3\\
 &  &  &  &  &  &  & 0.7\\
\bottomrule
\end{tabular}
\begin{tablenotes}
\item 
           T:C indicates Treatment:Control.
\end{tablenotes}
\end{threeparttable}
\end{table}

\begin{table}[!htbp]

\caption{\label{tab:setting2}Scenarios for simulating meta-analysis of proportions}
\centering
\begin{tabular}[t]{rrrrr}
\toprule
True log odds ($\theta$) & $\rho$ & $(P_{max}, P_{min})$ & Total subjects ($n_i$) & $\tau^2$\\
\midrule
-2 & 0.8 & (0.99, 0.20) & {}[15, 30] & 0.1\\
 &  &  &  & 0.3\\
 &  &  &  & 0.7\\
\addlinespace
 &  &  & {}[25, 100] & 0.1\\
 &  &  &  & 0.3\\
 &  &  &  & 0.7\\
\bottomrule
\end{tabular}
\end{table}

\begin{table}[!htbp]

\caption{\label{tab:set1}Averages of estimation bias of the lnOR ($\theta=-2$) among different models under the HN model based data-generating process. All estimates were multiplied by 100.}
\centering
\begin{threeparttable}
\begin{tabular}[t]{rrrrrrrrrrrrrrr}
\toprule
$S$ & Patients & T:C & $\tau^2$ & $N$ & NN$_P$ & HN$_P$ & CBN$_P$ & NN$_O$ & HN$_O$ & CBN$_O$ & CN (CP) & CS (CP) & HN$^\text{Prop}$ (CP) & CBN$^\text{Prop}$ (CP)\\
\midrule
15 & U[30, 60] & 1:1 & 0.1 & 10.6 & 18.0 & -2.7 & 39.7 & 24.3 & 3.3 & 39.4 & -7.4 (78.0) & -40.7 (66.4) & -7.7 (92.8) & 28.0 (80.2)\\
 &  &  & 0.3 & 10.6 & 21.9 & -1.3 & 40.3 & 33.6 & 11.8 & 46.8 & -4.7 (80.1) & -31.8 (72.2) & -1.6 (92.4) & 32.1 (81.2)\\
 &  &  & 0.7 & 10.6 & 28.9 & 2.2 & 41.4 & 45.7 & 22.0 & 53.9 & 3.7 (82.8) & -23.1 (76.9) & 7.9 (91.0) & 38.5 (78.4)\\
\addlinespace
 &  & 2:1 & 0.1 & 10.6 & 6.2 & -0.3 & 53.9 & 12.7 & 5.6 & 54.2 & -10.7 (80.8) & -50.5 (49.5) & -2.8 (93.4) & 44.0 (61.0)\\
 &  &  & 0.3 & 10.5 & 6.5 & -0.9 & 52.6 & 18.8 & 12.3 & 57.2 & -7.7 (79.3) & -47.5 (52.0) & 0.3 (92.4) & 44.9 (64.6)\\
 &  &  & 0.7 & 10.5 & 9.8 & -0.7 & 52.2 & 27.7 & 19.1 & 61.2 & -3.5 (81.2) & -39.9 (53.3) & 4.3 (91.8) & 45.7 (67.1)\\
\addlinespace
 & U[50, 200] & 1:1 & 0.1 & 10.7 & 25.3 & -4.8 & 10.7 & 31.2 & 3.3 & 13.4 & -2.2 (76.3) & -23.6 (86.2) & -7.0 (91.8) & 2.5 (91.6)\\
 &  &  & 0.3 & 10.8 & 30.3 & -4.9 & 10.2 & 40.2 & 8.4 & 19.5 & -1.5 (79.2) & -14.8 (88.7) & -4.4 (91.4) & 6.1 (90.8)\\
 &  &  & 0.7 & 10.8 & 40.0 & 1.3 & 15.4 & 54.9 & 21.8 & 31.7 & 9.4 (81.2) & -2.5 (86.5) & 9.0 (88.8) & 18.3 (87.1)\\
\addlinespace
 &  & 2:1 & 0.1 & 10.9 & 11.5 & -3.6 & 16.6 & 17.7 & 2.8 & 19.5 & -11.7 (79.6) & -41.6 (54.3) & -6.4 (92.1) & 9.1 (90.6)\\
 &  &  & 0.3 & 10.8 & 16.4 & -1.4 & 19.0 & 27.5 & 11.8 & 26.6 & -5.9 (78.2) & -33.5 (57.4) & 1.5 (92.1) & 14.1 (87.9)\\
 &  &  & 0.7 & 10.8 & 21.3 & 0.4 & 19.3 & 37.4 & 19.3 & 33.3 & -0.1 (80.9) & -25.6 (63.3) & 6.0 (91.8) & 18.5 (86.9)\\
\addlinespace
 & U[500, 700] & 1:1 & 0.1 & 10.4 & 29.6 & -6.5 & -5.7 & 34.6 & 0.4 & 2.8 & -1.0 (75.3) & -17.0 (94.8) & -11.0 (89.9) & -8.1 (90.0)\\
 &  &  & 0.3 & 10.4 & 36.4 & -2.7 & 0.2 & 45.7 & 11.7 & 14.3 & 2.9 (80.2) & -7.7 (91.3) & -1.5 (91.9) & 1.2 (92.1)\\
 &  &  & 0.7 & 10.4 & 46.2 & 3.4 & 5.8 & 59.7 & 24.1 & 25.9 & 11.2 (80.2) & 5.1 (89.1) & 9.8 (89.3) & 11.7 (89.3)\\
\addlinespace
 &  & 2:1 & 0.1 & 10.3 & 17.5 & -2.9 & 0.4 & 23.1 & 4.1 & 7.9 & -10.6 (76.9) & -31.9 (69.0) & -6.4 (92.5) & -2.6 (92.8)\\
 &  &  & 0.3 & 10.4 & 22.0 & -1.3 & 2.1 & 31.6 & 11.4 & 14.7 & -7.6 (76.5) & -26.2 (66.1) & 0.0 (91.1) & 2.3 (91.2)\\
 &  &  & 0.7 & 10.4 & 28.8 & 1.7 & 5.0 & 44.4 & 23.5 & 25.7 & 3.0 (81.1) & -12.4 (75.9) & 8.4 (91.4) & 11.6 (89.8)\\
\addlinespace
50 & U[30, 60] & 1:1 & 0.1 & 35.8 & 18.4 & -2.4 & 40.9 & 25.7 & 5.1 & 43.0 & -10.7 (87.4) & -31.6 (41.7) & -4.3 (93.2) & 32.9 (58.7)\\
 &  &  & 0.3 & 35.9 & 23.5 & -1.4 & 41.6 & 35.5 & 12.6 & 48.5 & -7.4 (89.0) & -22.7 (59.7) & -0.2 (93.3) & 35.6 (57.5)\\
 &  &  & 0.7 & 36.0 & 30.4 & 1.3 & 42.7 & 48.0 & 22.4 & 56.1 & 3.5 (91.3) & -11.5 (76.8) & 5.7 (94.2) & 38.5 (56.0)\\
\addlinespace
 &  & 2:1 & 0.1 & 35.9 & 6.2 & -1.1 & 54.1 & 13.2 & 6.5 & 55.3 & -12.3 (87.4) & -62.2 (11.1) & -1.0 (92.3) & 46.6 (12.6)\\
 &  &  & 0.3 & 35.9 & 9.1 & -0.0 & 55.0 & 21.2 & 13.4 & 59.9 & -8.9 (90.9) & -58.4 (9.6) & 1.8 (94.7) & 48.1 (14.5)\\
 &  &  & 0.7 & 35.8 & 11.3 & 0.2 & 54.9 & 30.6 & 21.8 & 65.3 & 0.7 (90.7) & -52.0 (17.1) & 5.3 (93.0) & 48.6 (24.5)\\
\addlinespace
 & U[50, 200] & 1:1 & 0.1 & 36.9 & 27.3 & -2.8 & 13.2 & 33.6 & 4.9 & 17.5 & -6.2 (89.0) & -9.3 (89.0) & -3.7 (93.4) & 7.5 (91.9)\\
 &  &  & 0.3 & 37.1 & 34.4 & -0.3 & 15.2 & 44.2 & 12.7 & 25.1 & -2.0 (90.6) & 1.0 (91.8) & 1.3 (93.2) & 12.2 (89.9)\\
 &  &  & 0.7 & 37.0 & 42.2 & 1.1 & 15.7 & 56.9 & 20.8 & 32.1 & 6.4 (92.3) & 13.9 (75.5) & 5.3 (94.2) & 15.1 (90.3)\\
\addlinespace
 &  & 2:1 & 0.1 & 36.9 & 13.3 & -1.7 & 19.2 & 20.3 & 6.4 & 22.5 & -12.4 (84.1) & -40.6 (9.8) & -2.1 (93.8) & 13.0 (85.0)\\
 &  &  & 0.3 & 37.0 & 18.9 & 0.2 & 20.2 & 30.0 & 13.2 & 28.8 & -7.5 (89.7) & -34.1 (18.3) & 2.9 (94.3) & 16.4 (82.9)\\
 &  &  & 0.7 & 37.1 & 23.4 & 1.6 & 20.8 & 39.9 & 21.2 & 35.7 & 1.1 (91.3) & -27.1 (33.0) & 4.6 (93.3) & 17.2 (85.0)\\
\addlinespace
 & U[500, 700] & 1:1 & 0.1 & 35.4 & 31.3 & -3.6 & -1.1 & 37.0 & 4.0 & 7.0 & -5.4 (88.0) & -3.1 (94.4) & -5.9 (94.3) & -3.2 (94.9)\\
 &  &  & 0.3 & 35.4 & 38.9 & -1.1 & 1.5 & 48.1 & 12.8 & 15.2 & -1.4 (90.4) & 8.6 (88.4) & 0.6 (94.0) & 3.2 (93.5)\\
 &  &  & 0.7 & 35.3 & 47.6 & 1.2 & 3.6 & 62.2 & 23.0 & 25.2 & 7.4 (92.2) & 23.9 (58.8) & 5.4 (94.2) & 7.7 (93.3)\\
\addlinespace
 &  & 2:1 & 0.1 & 35.3 & 19.1 & -0.7 & 2.7 & 25.6 & 7.1 & 10.6 & -10.5 (87.0) & -27.7 (36.9) & -1.2 (93.2) & 2.4 (92.7)\\
 &  &  & 0.3 & 35.3 & 24.6 & 0.5 & 3.9 & 35.1 & 15.1 & 18.1 & -7.3 (88.6) & -20.8 (50.9) & 2.8 (94.0) & 6.1 (92.5)\\
 &  &  & 0.7 & 35.4 & 29.8 & 0.9 & 4.1 & 45.8 & 23.1 & 25.9 & 2.2 (90.7) & -12.7 (72.4) & 5.5 (93.3) & 8.6 (92.0)\\
\bottomrule
\end{tabular}
\begin{tablenotes}
\item 
           Results of NN$_P$, HN$_P$, and CBN$_P$ are based on the population studies;
           results of NN$_O$, HN$_O$, and CBN$_O$ are based on the published studies;
           CN and CS indicate the Copas-N and Copas-Shi methods;
           HN$^\text{Prop}$ and CBN$^\text{Prop}$ indicate the proposed HN or CBN model based sensitivity analysis methods;
           CP indicates the coverage probability.
\end{tablenotes}
\end{threeparttable}
\end{table}

\begin{table}[!htbp]

\caption{\label{tab:set2}Averages of estimation bias of the lnOR ($\theta=-2$) among different models under the 2SBN model based data-generating process. All estimates were multiplied by 100.}
\centering
\begin{threeparttable}
\begin{tabular}[t]{rrrrrrrrrrrrrrr}
\toprule
$S$ & Patients & T:C & $\tau^2$ & $N$ & NN$_P$ & HN$_P$ & CBN$_P$ & NN$_O$ & HN$_O$ & CBN$_O$ & CN (CP) & CS (CP) & HN$^\text{Prop}$ (CP) & CBN$^\text{Prop}$ (CP)\\
\midrule
15 & U[30, 60] & 1:1 & 0.1 & 10.6 & 66.7 & -7.8 & 6.8 & 66.7 & -4.5 & 13.6 & 69.8 (48.2) & 60.3 (93.2) & -7.3 (92.6) & 6.9 (92.7)\\
 &  &  & 0.3 & 10.6 & 69.2 & -4.1 & 12.1 & 70.9 & 2.0 & 18.5 & 70.6 (46.5) & 62.4 (90.3) & -2.5 (93.0) & 12.6 (89.6)\\
 &  &  & 0.7 & 10.5 & 79.0 & 10.8 & 25.1 & 83.6 & 23.1 & 38.4 & 77.5 (40.5) & 65.7 (86.6) & 16.3 (89.2) & 31.7 (87.0)\\
\addlinespace
 &  & 2:1 & 0.1 & 10.6 & 42.4 & -3.7 & 8.1 & 43.6 & -1.6 & 14.0 & 42.3 (66.7) & 41.4 (93.0) & -10.9 (93.8) & 4.2 (91.4)\\
 &  &  & 0.3 & 10.6 & 48.2 & 3.0 & 18.9 & 52.0 & 11.6 & 27.0 & 48.3 (65.5) & 42.0 (94.3) & 6.5 (91.6) & 20.4 (89.0)\\
 &  &  & 0.7 & 10.6 & 54.2 & 11.2 & 27.2 & 61.0 & 25.9 & 40.2 & 52.5 (60.7) & 43.6 (91.8) & 16.8 (90.8) & 31.1 (86.6)\\
\addlinespace
 & U[50, 200] & 1:1 & 0.1 & 10.8 & 34.1 & -2.2 & 11.9 & 34.6 & 0.8 & 16.8 & 34.1 (76.9) & 5.9 (90.9) & -7.1 (90.7) & 8.3 (90.0)\\
 &  &  & 0.3 & 10.8 & 41.0 & 0.4 & 16.2 & 43.9 & 8.8 & 24.3 & 38.6 (75.3) & 12.9 (88.9) & 1.7 (90.2) & 17.0 (86.7)\\
 &  &  & 0.7 & 10.9 & 52.0 & 8.7 & 25.0 & 57.7 & 23.9 & 38.4 & 45.3 (67.5) & 22.6 (80.5) & 14.9 (88.3) & 29.3 (79.9)\\
\addlinespace
 &  & 2:1 & 0.1 & 10.8 & 22.8 & -0.1 & 15.4 & 24.6 & 5.0 & 20.8 & 20.6 (78.4) & 1.9 (81.2) & -2.2 (91.7) & 13.7 (85.3)\\
 &  &  & 0.3 & 10.9 & 29.9 & 5.1 & 20.7 & 34.4 & 13.9 & 30.6 & 25.0 (78.2) & 5.5 (81.8) & 7.4 (92.0) & 21.7 (85.6)\\
 &  &  & 0.7 & 10.8 & 40.0 & 10.0 & 25.2 & 49.0 & 27.4 & 41.9 & 32.7 (75.1) & 15.0 (80.6) & 16.7 (87.9) & 31.9 (79.6)\\
\addlinespace
 & U[500, 700] & 1:1 & 0.1 & 10.5 & 10.9 & 0.7 & 15.6 & 17.2 & 8.4 & 24.3 & 24.3 (77.1) & -11.5 (61.4) & 1.5 (92.1) & 17.2 (77.7)\\
 &  &  & 0.3 & 10.5 & 15.8 & 1.5 & 16.7 & 26.8 & 15.6 & 30.2 & 28.0 (76.4) & -5.4 (72.1) & 7.1 (88.5) & 22.2 (77.4)\\
 &  &  & 0.7 & 10.5 & 22.7 & 3.8 & 18.3 & 39.2 & 24.5 & 38.4 & 27.3 (75.9) & 9.9 (78.0) & 9.9 (85.9) & 24.3 (78.9)\\
\addlinespace
 &  & 2:1 & 0.1 & 10.4 & 7.2 & 0.8 & 16.1 & 13.7 & 8.5 & 23.2 & 15.5 (74.8) & -11.7 (61.0) & 0.8 (89.3) & 16.8 (75.3)\\
 &  &  & 0.3 & 10.4 & 12.2 & 2.9 & 17.7 & 23.6 & 16.7 & 31.1 & 17.0 (75.4) & -7.0 (72.8) & 6.6 (86.8) & 21.0 (76.7)\\
 &  &  & 0.7 & 10.4 & 14.7 & 1.5 & 16.2 & 31.8 & 22.3 & 36.3 & 18.7 (77.4) & 2.0 (80.5) & 8.1 (85.5) & 22.2 (78.4)\\
\addlinespace
50 & U[30, 60] & 1:1 & 0.1 & 36.0 & 72.3 & -5.5 & 10.3 & 71.8 & 0.9 & 14.2 & 78.7 (5.5) & 65.1 (53.0) & -10.1 (90.3) & 7.3 (90.1)\\
 &  &  & 0.3 & 35.9 & 76.4 & -0.2 & 15.7 & 78.4 & 10.6 & 26.0 & 81.9 (3.2) & 67.4 (43.2) & -1.2 (92.6) & 15.8 (88.7)\\
 &  &  & 0.7 & 35.9 & 84.8 & 10.6 & 27.0 & 89.6 & 26.2 & 43.0 & 87.8 (1.0) & 71.0 (28.7) & 13.8 (91.4) & 30.3 (81.8)\\
\addlinespace
 &  & 2:1 & 0.1 & 36.1 & 47.7 & -2.7 & 14.1 & 49.1 & 3.6 & 19.9 & 52.4 (31.0) & 44.3 (83.9) & -4.0 (91.6) & 12.7 (87.4)\\
 &  &  & 0.3 & 36.0 & 52.6 & 4.1 & 21.0 & 56.3 & 15.2 & 32.4 & 55.3 (25.4) & 46.0 (80.6) & 5.7 (92.8) & 22.1 (85.8)\\
 &  &  & 0.7 & 35.9 & 62.0 & 14.4 & 31.2 & 68.4 & 30.3 & 46.2 & 60.9 (18.4) & 49.3 (69.2) & 17.6 (89.0) & 34.3 (80.2)\\
\addlinespace
 & U[50, 200] & 1:1 & 0.1 & 37.2 & 39.4 & -1.3 & 15.5 & 40.1 & 4.8 & 20.9 & 48.0 (33.5) & 15.8 (94.5) & -0.1 (93.5) & 16.0 (82.1)\\
 &  &  & 0.3 & 37.1 & 47.7 & 2.9 & 19.6 & 50.6 & 13.0 & 28.6 & 52.8 (25.4) & 19.6 (83.3) & 5.5 (92.1) & 22.0 (78.4)\\
 &  &  & 0.7 & 37.0 & 58.4 & 7.3 & 23.4 & 64.4 & 23.1 & 38.7 & 59.4 (14.4) & 31.7 (58.7) & 11.9 (88.8) & 27.7 (75.3)\\
\addlinespace
 &  & 2:1 & 0.1 & 37.0 & 26.3 & 0.5 & 16.8 & 27.9 & 6.1 & 21.8 & 29.5 (72.4) & 3.5 (89.5) & 1.8 (94.1) & 18.3 (80.0)\\
 &  &  & 0.3 & 37.0 & 32.7 & 3.5 & 20.1 & 36.9 & 13.7 & 29.4 & 32.4 (67.2) & 9.2 (90.7) & 6.3 (92.2) & 22.9 (77.3)\\
 &  &  & 0.7 & 37.1 & 44.3 & 9.9 & 25.6 & 51.8 & 25.5 & 40.7 & 39.4 (51.5) & 18.8 (81.5) & 13.2 (88.4) & 28.9 (75.7)\\
\addlinespace
 & U[500, 700] & 1:1 & 0.1 & 35.2 & 12.4 & 0.4 & 15.7 & 18.4 & 8.4 & 23.2 & 34.0 (62.8) & -14.4 (51.2) & 1.5 (93.9) & 16.9 (69.2)\\
 &  &  & 0.3 & 35.3 & 18.1 & 1.4 & 16.4 & 29.0 & 15.7 & 30.4 & 36.2 (61.2) & -1.0 (83.8) & 4.3 (90.7) & 19.4 (71.2)\\
 &  &  & 0.7 & 35.3 & 26.0 & 3.1 & 17.7 & 42.4 & 25.2 & 39.1 & 42.3 (50.7) & 18.6 (67.3) & 6.5 (90.5) & 21.5 (79.1)\\
\addlinespace
 &  & 2:1 & 0.1 & 35.3 & 8.6 & 1.3 & 16.5 & 15.2 & 9.0 & 24.0 & 23.0 (71.7) & -14.9 (52.5) & 2.6 (92.5) & 17.9 (64.2)\\
 &  &  & 0.3 & 35.3 & 13.1 & 2.7 & 17.5 & 24.6 & 16.8 & 31.2 & 23.1 (76.2) & -5.3 (83.2) & 4.4 (91.2) & 19.6 (71.1)\\
 &  &  & 0.7 & 35.4 & 19.0 & 3.7 & 18.0 & 36.1 & 25.3 & 38.9 & 26.5 (73.3) & 10.5 (81.1) & 5.9 (90.1) & 20.5 (78.0)\\
\bottomrule
\end{tabular}
\begin{tablenotes}
\item 
           Results of NN$_P$, HN$_P$, and CBN$_P$ are based on the population studies;
           results of NN$_O$, HN$_O$, and CBN$_O$ are based on the published studies;
           CN and CS indicate the Copas-N and Copas-Shi methods;
           HN$^\text{Prop}$ and CBN$^\text{Prop}$ indicate the proposed HN or CBN model based sensitivity analysis methods;
           CP indicates the coverage probability.
\end{tablenotes}
\end{threeparttable}
\end{table}

\begin{table}[!htbp]

\caption{\label{tab:set3}Averages of estimation bias of the log odds ($\theta=-2$) among different models under the 1SBN model based data-generating process. All estimates were multiplied by 100.}
\centering
\begin{threeparttable}
\begin{tabular}[t]{rrrrrrrrrrr}
\toprule
$S$ & Patients & $\tau^2$ & $N$ & NN$_P$ & 1SBN$_P$ & NN$_O$ & 1SBN$_O$ & CN (CP) & CS (CP) & 1SBN$^{\text{prop}}$ (CP)\\
\midrule
15 & U[15, 30] & 0.1 & 10.6 & 18.7 & -1.3 & 22.4 & 4.8 & -13.2 (78.6) & -18.9 (74.4) & -1.5 (93.3)\\
 &  & 0.3 & 10.6 & 24.6 & -0.0 & 32.2 & 12.1 & -10.9 (78.8) & -5.8 (79.7) & 1.4 (91.1)\\
 &  & 0.7 & 10.6 & 29.7 & 0.7 & 42.4 & 19.8 & -3.9 (80.3) & 9.6 (73.6) & 7.5 (90.8)\\
\addlinespace
 & U[25, 100] & 0.1 & 10.8 & 7.6 & -1.0 & 11.8 & 5.6 & -4.7 (80.0) & -13.1 (64.0) & 0.4 (93.0)\\
 &  & 0.3 & 10.8 & 10.4 & -0.1 & 18.1 & 11.1 & -3.3 (79.4) & -5.0 (69.2) & 2.8 (90.1)\\
 &  & 0.7 & 10.8 & 12.9 & -0.2 & 26.3 & 17.8 & -2.4 (80.1) & 2.5 (73.6) & 4.3 (87.0)\\
\addlinespace
50 & U[15, 30] & 0.1 & 36.0 & 20.6 & -0.4 & 25.0 & 6.8 & -13.4 (86.1) & -3.8 (82.3) & 0.3 (94.8)\\
 &  & 0.3 & 36.1 & 25.4 & 0.1 & 33.9 & 12.9 & -11.5 (86.6) & 9.4 (75.6) & 2.7 (92.2)\\
 &  & 0.7 & 36.2 & 30.1 & 0.6 & 43.6 & 20.4 & -5.3 (90.7) & 22.9 (35.9) & 4.0 (92.5)\\
\addlinespace
 & U[25, 100] & 0.1 & 37.0 & 8.7 & 0.3 & 13.0 & 6.6 & -2.5 (93.4) & -11.3 (43.0) & 1.9 (92.1)\\
 &  & 0.3 & 37.1 & 10.2 & -0.3 & 18.6 & 11.4 & -3.0 (92.5) & -3.3 (75.1) & 1.5 (93.2)\\
 &  & 0.7 & 37.0 & 13.2 & 0.2 & 26.9 & 18.7 & -2.4 (92.7) & 6.0 (82.5) & 1.4 (93.0)\\
\bottomrule
\end{tabular}
\begin{tablenotes}
\item 
           NN$_P$ and BN$_P$ are the estimates based on the population studies;
           NN$_O$ and BN$_O$ are the estimates based on the published studies;
           CN and CS are the Copas-N and Copas-Shi methods;
           1SBN$^{\text{prop}}$ are the proposed 1SBN model based sensitivity analysis methods;
           CP indicates the coverage probability.
\end{tablenotes}
\end{threeparttable}
\end{table}

\clearpage
\appendix

\clearpage
\section{Review of the Copas-N method for publication bias in the normal-normal random effects model}


As mentioned in Section 4 of the main text, we review the Copas-N sensitivity analysis method\cite{Copas1999}.

Copas\cite{Copas1999} extended the Heckman-type selection function and proposed a selection-model-based method to address PB on associations between treatment and the continuous outcomes.
For binary data, the outcomes can be the log-transformed odds ratios (ORs) or the log-transformed odds.
Then, the outcomes were re-expressed by the following random-effects model: 
\begin{align}\label{yi}
\hat\thetai = \theta+{(\Ni^{-1}+\tau^2)}^{1/2}\epsilon_i,  
\end{align}
where $\epsilon_i$ is the random residual.
However, the likelihood \eqref{yi} is not exactly same with the NN model.
To address PB, Copas\cite{Copas1999} introduced a latent Gaussian variable $Z_i$ with the selection equation defined as:
\begin{align*}
Z_i = \gamma_0+\gamma_1\Ni^{1/2}+\delta_i.
\end{align*}
In the above equations, $\epsilon_i$ and $\delta_i$ were regarded as random residuals with the following distribution:
\begin{align*}
\binom{\epsilon_i}{\delta_i}&\sim N\left( \binom{0}{0}, \smat{1}{\rho}{\rho}{1}\right),
\end{align*}
where $\gamma_0$ and $\gamma_1$ are constants and $\rho$ indicates the correlation between $\epsilon_i$ and $\delta_i$.
Under selective publication of studies, the log-likelihood conditional on the published studies ($Z_i>0$) was derived in equation (7) of Copas,\cite{Copas1999} shown as follows:
\begin{align}
\ell(\theta,\tau,\rho,\gamma_0,\gamma_1)=\sumi \left\{-\dfrac{1}{2}\log\left(\tau^2+\dfrac{1}{n_i}\right)-\dfrac{(y-\theta)^2}{2(\tau^2+n_i^{-1})}-\log\Phi(\gamma_0+\gamma_1\Ni^{1/2}) + \log\Phi(v_i) \right\}
\end{align}
with
\begin{align*}
v_i = \dfrac{\gamma_0+\gamma_1\Ni^{1/2} + \rho(\hat\thetai-\theta)/\sqrt{\tau^2+n_i^{-1}}}{\sqrt{1-\rho^2}}
\end{align*}
Given fixed pairs of values for $(\gamma_0,\gamma_1)$, the rest parameters can be estimated by the maximum likelihood estimation.

\clearpage
\section{Additional information in simulation studies}\label{sec:rfuncs}

\subsection{Implementations}
Simulation studies were conducted by R (version 4.0.3) on system of Intel Xeon Platinum 8368.
The random numbers from hypergeometric distribution (equation 7 in the main text) were generated by the R function \texttt{rFNCHypergeo()} in the R package \texttt{BiasedUrn}.\cite{Fog2008a,Fog2008}
The random numbers from binomial distribution were generated by \texttt{rbinom()} in the R package \texttt{stats}.
The maximum likelihood estimations were optimized by the numerical Newton-Raphson method and implemented by \texttt{nlminb()} in the R package \texttt{stats} with initial values set as true parameters plus the random values from uniform distribution $U[-0.1,0.1]$.
The settings of initial value might be arbitrary.
The also implemented different initial values by adopting the estimated $\hat\theta$ from the normal-normal (NN) models as initial values for $\theta$ and $0.5$ for $\tau$ for estimation; however, the estimates had only small changes and the conclusions did not change.
The integrations in the hypergeometric-normal (HN) and binomial-normal (BN) models were implemented numerically by \texttt{hcubature()} in the R package \texttt{cubature}.\cite{cubature}

\subsection{Summary of the amount of rare events}

As mentioned in Section 5.2 of the main text, we summarized the proportion of studies with less-than-three events in simulation studies.
Corresponding to different data-generating processes and scenarios of parameters, for meta-analysis of ORs, the proportions of studies with rare events in treatment and control group were summarized and presented in Table \ref{tab:prop1}.
For meta-analysis of proportions,  the proportions of studies with rare events were summarized and presented in Table \ref{tab:prop2}.
The subscripts $P$ and $O$ indicates the proportions of rare events in the population studies (published and unpublished studies) and the published (observed) studies, respectively.

\begin{table}[!htbp]

\caption{\label{tab:prop1}Summary of the proportions of studies with rare events under the HN and 2SBN model based data-generating processes.}
\centering
\begin{threeparttable}
\begin{tabular}[t]{rrrrrrrr}
\toprule
S & Patients & T:C & $\tau^2$ & HN$_P$ & HN$_O$ & 2SBN$_P$ & 2SBN$_O$\\
\midrule
15 & U[30, 60] & 1:1 & 0.1 & 89.1 & 88.7 & 99.6 & 99.4\\
 &  &  & 0.3 & 87.8 & 86.2 & 99.2 & 98.8\\
 &  &  & 0.7 & 85.2 & 82.6 & 97.8 & 96.9\\
\addlinespace
 &  & 2:1 & 0.1 & 61.5 & 61.7 & 99.5 & 99.3\\
 &  &  & 0.3 & 61.6 & 60.8 & 98.8 & 98.3\\
 &  &  & 0.7 & 63.5 & 61.7 & 97.5 & 96.6\\
\addlinespace
 & U[50, 200] & 1:1 & 0.1 & 94.6 & 93.7 & 91.1 & 88.3\\
 &  &  & 0.3 & 92.3 & 90.8 & 88.3 & 84.5\\
 &  &  & 0.7 & 89.2 & 86.5 & 84.7 & 79.9\\
\addlinespace
 &  & 2:1 & 0.1 & 76.0 & 75.2 & 85.0 & 80.4\\
 &  &  & 0.3 & 74.7 & 72.4 & 83.2 & 78.0\\
 &  &  & 0.7 & 74.6 & 71.8 & 81.0 & 74.9\\
\addlinespace
 & U[500, 700] & 1:1 & 0.1 & 95.6 & 94.6 & 36.5 & 32.3\\
 &  &  & 0.3 & 93.6 & 91.9 & 39.8 & 34.3\\
 &  &  & 0.7 & 90.6 & 87.9 & 43.3 & 36.6\\
\addlinespace
 &  & 2:1 & 0.1 & 81.3 & 79.0 & 29.1 & 26.0\\
 &  &  & 0.3 & 80.1 & 76.6 & 32.3 & 27.7\\
 &  &  & 0.7 & 78.4 & 74.2 & 37.6 & 31.6\\
\addlinespace
50 & U[30, 60] & 1:1 & 0.1 & 89.5 & 88.9 & 99.7 & 99.6\\
 &  &  & 0.3 & 87.7 & 86.1 & 99.2 & 98.9\\
 &  &  & 0.7 & 84.7 & 82.0 & 98.0 & 97.2\\
\addlinespace
 &  & 2:1 & 0.1 & 61.5 & 61.4 & 99.5 & 99.4\\
 &  &  & 0.3 & 62.2 & 60.9 & 98.9 & 98.5\\
 &  &  & 0.7 & 63.5 & 61.7 & 97.7 & 96.8\\
\addlinespace
 & U[50, 200] & 1:1 & 0.1 & 94.3 & 93.7 & 91.7 & 89.3\\
 &  &  & 0.3 & 92.1 & 90.7 & 89.3 & 86.0\\
 &  &  & 0.7 & 89.1 & 86.7 & 86.2 & 81.8\\
\addlinespace
 &  & 2:1 & 0.1 & 76.2 & 75.0 & 86.6 & 82.9\\
 &  &  & 0.3 & 74.9 & 72.6 & 84.9 & 80.5\\
 &  &  & 0.7 & 74.4 & 71.3 & 82.5 & 77.3\\
\addlinespace
 & U[500, 700] & 1:1 & 0.1 & 95.7 & 95.0 & 43.0 & 39.2\\
 &  &  & 0.3 & 93.8 & 92.3 & 45.9 & 40.9\\
 &  &  & 0.7 & 90.7 & 88.0 & 49.0 & 42.9\\
\addlinespace
 &  & 2:1 & 0.1 & 81.7 & 79.3 & 36.1 & 33.7\\
 &  &  & 0.3 & 79.9 & 76.2 & 39.0 & 35.1\\
 &  &  & 0.7 & 78.6 & 74.3 & 43.4 & 38.2\\
\bottomrule
\end{tabular}
\begin{tablenotes}
\item 
           Results of HN$_P$ and 2SBN$_P$ are based on the population studies;
           results of HN$_O$ and 2SBN$_O$ are based on the published studies.
\end{tablenotes}
\end{threeparttable}
\end{table}

\begin{table}[!htbp]

\caption{\label{tab:prop2}Summary of the proportions of studies with rare events under the 1SBN model based data-generating processes.}
\centering
\begin{threeparttable}
\begin{tabular}[t]{rrrrr}
\toprule
S & Patients & $\tau^2$ & 1SBN$_P$ & 1SBN$_O$\\
\midrule
15 & U[15, 30] & 0.1 & 91.9 & 89.0\\
 &  & 0.3 & 87.2 & 82.4\\
 &  & 0.7 & 81.6 & 75.1\\
\addlinespace
 & U[25, 100] & 0.1 & 34.9 & 20.1\\
 &  & 0.3 & 37.6 & 22.1\\
 &  & 0.7 & 40.4 & 25.0\\
\addlinespace
50 & U[15, 30] & 0.1 & 91.9 & 89.3\\
 &  & 0.3 & 87.6 & 83.2\\
 &  & 0.7 & 81.6 & 75.3\\
\addlinespace
 & U[25, 100] & 0.1 & 34.7 & 20.9\\
 &  & 0.3 & 37.5 & 22.8\\
 &  & 0.7 & 40.1 & 25.6\\
\bottomrule
\end{tabular}
\begin{tablenotes}
\item 
           Results of 1SBN$_P$ are based on the population studies;
           results of 1SBN$_O$ are based on the published studies.
\end{tablenotes}
\end{threeparttable}
\end{table}

\subsection{Additional result 1: meta-analysis of odds ratios under the HN model based data-generating process}

As mentioned in Sections 5.2, we presented convergence rates (Table \ref{tab:set1}) and average estimates of $\tau^2$ (Table \ref{tab:set1-tau}) when population data were simulated under the HN model based data-generating process.


\begin{table}[!htbp]

\caption{\label{tab:set1}Convergence proportion (\%) of estimations among different models under the HN model based data-generating process.}
\centering
\begin{threeparttable}
\begin{tabular}[t]{rrrrrrrrrrrrrrr}
\toprule
$S$ & Patients & T:C & $\tau^2$ & $N$ & NN$_P$ & HN$_P$ & CBN$_P$ & NN$_O$ & HN$_O$ & CBN$_O$ & CN & CS & HN$^\text{Prop}$ & CBN$^\text{Prop}$\\
\midrule
15 & U[30, 60] & 1:1 & 0.1 & 10.6 & 100.0 & 71.6 & 63.2 & 100.0 & 69.6 & 66.8 & 100 & 100.0 & 61.5 & 75.2\\
 &  &  & 0.3 & 10.6 & 99.7 & 80.0 & 73.6 & 99.9 & 75.5 & 74.1 & 100 & 99.9 & 72.2 & 82.3\\
 &  &  & 0.7 & 10.6 & 99.2 & 92.3 & 86.2 & 100.0 & 88.6 & 83.6 & 100 & 100.0 & 85.2 & 90.2\\
\addlinespace
 &  & 2:1 & 0.1 & 10.6 & 100.0 & 74.9 & 66.1 & 100.0 & 73.4 & 67.9 & 100 & 100.0 & 62.4 & 80.9\\
 &  &  & 0.3 & 10.5 & 99.9 & 86.3 & 73.3 & 99.7 & 82.2 & 74.4 & 100 & 99.9 & 73.9 & 84.9\\
 &  &  & 0.7 & 10.5 & 99.2 & 95.4 & 87.2 & 99.8 & 90.8 & 84.2 & 100 & 100.0 & 86.3 & 91.2\\
\addlinespace
 & U[50, 200] & 1:1 & 0.1 & 10.7 & 99.9 & 69.8 & 68.0 & 99.8 & 69.3 & 66.4 & 100 & 100.0 & 57.9 & 70.2\\
 &  &  & 0.3 & 10.8 & 99.3 & 81.9 & 80.4 & 99.9 & 79.1 & 76.6 & 100 & 100.0 & 73.3 & 81.6\\
 &  &  & 0.7 & 10.8 & 99.2 & 92.5 & 89.6 & 99.5 & 88.5 & 87.2 & 100 & 100.0 & 86.0 & 91.3\\
\addlinespace
 &  & 2:1 & 0.1 & 10.9 & 100.0 & 74.7 & 69.3 & 100.0 & 73.2 & 70.2 & 100 & 100.0 & 65.9 & 76.7\\
 &  &  & 0.3 & 10.8 & 99.6 & 86.7 & 83.2 & 99.8 & 82.8 & 79.4 & 100 & 100.0 & 78.0 & 86.0\\
 &  &  & 0.7 & 10.8 & 99.5 & 95.5 & 94.3 & 99.5 & 90.8 & 91.4 & 100 & 100.0 & 88.6 & 94.3\\
\addlinespace
 & U[500, 700] & 1:1 & 0.1 & 10.4 & 99.9 & 73.4 & 71.1 & 99.8 & 68.6 & 69.6 & 100 & 100.0 & 60.5 & 70.1\\
 &  &  & 0.3 & 10.4 & 100.0 & 79.8 & 81.8 & 99.6 & 74.2 & 75.7 & 100 & 100.0 & 69.6 & 79.2\\
 &  &  & 0.7 & 10.4 & 99.5 & 89.9 & 89.7 & 99.9 & 86.7 & 85.7 & 100 & 100.0 & 85.1 & 92.2\\
\addlinespace
 &  & 2:1 & 0.1 & 10.3 & 99.8 & 77.8 & 75.4 & 100.0 & 72.9 & 73.5 & 100 & 100.0 & 65.0 & 73.7\\
 &  &  & 0.3 & 10.4 & 99.6 & 85.4 & 85.7 & 99.5 & 81.4 & 81.4 & 100 & 100.0 & 79.2 & 84.6\\
 &  &  & 0.7 & 10.4 & 99.7 & 95.0 & 95.4 & 99.7 & 92.1 & 91.0 & 100 & 100.0 & 88.7 & 93.9\\
\addlinespace
50 & U[30, 60] & 1:1 & 0.1 & 35.8 & 99.8 & 80.6 & 67.0 & 99.5 & 77.4 & 69.3 & 100 & 100.0 & 73.6 & 86.7\\
 &  &  & 0.3 & 35.9 & 100.0 & 92.4 & 83.0 & 100.0 & 89.3 & 80.9 & 100 & 100.0 & 88.7 & 94.0\\
 &  &  & 0.7 & 36.0 & 99.6 & 99.4 & 97.3 & 99.9 & 97.4 & 94.8 & 100 & 100.0 & 97.3 & 97.8\\
\addlinespace
 &  & 2:1 & 0.1 & 35.9 & 99.8 & 81.5 & 63.3 & 99.8 & 76.5 & 65.3 & 100 & 100.0 & 76.1 & 89.4\\
 &  &  & 0.3 & 35.9 & 99.9 & 96.4 & 81.6 & 100.0 & 92.8 & 81.5 & 100 & 100.0 & 91.8 & 96.2\\
 &  &  & 0.7 & 35.8 & 99.8 & 100.0 & 97.2 & 99.7 & 99.6 & 95.6 & 100 & 100.0 & 98.4 & 99.0\\
\addlinespace
 & U[50, 200] & 1:1 & 0.1 & 36.9 & 99.2 & 80.0 & 74.1 & 99.6 & 77.7 & 74.5 & 100 & 100.0 & 75.1 & 82.5\\
 &  &  & 0.3 & 37.1 & 99.5 & 92.5 & 89.1 & 99.3 & 88.4 & 86.1 & 100 & 100.0 & 89.1 & 93.3\\
 &  &  & 0.7 & 37.0 & 99.9 & 99.6 & 98.9 & 99.6 & 98.7 & 97.2 & 100 & 100.0 & 97.0 & 98.4\\
\addlinespace
 &  & 2:1 & 0.1 & 36.9 & 99.6 & 82.9 & 75.9 & 100.0 & 79.8 & 75.3 & 100 & 100.0 & 77.6 & 88.8\\
 &  &  & 0.3 & 37.0 & 99.6 & 97.2 & 94.3 & 99.8 & 93.9 & 92.5 & 100 & 100.0 & 94.4 & 96.4\\
 &  &  & 0.7 & 37.1 & 99.7 & 100.0 & 99.8 & 99.8 & 99.6 & 99.5 & 100 & 99.9 & 98.4 & 99.9\\
\addlinespace
 & U[500, 700] & 1:1 & 0.1 & 35.4 & 99.1 & 76.6 & 74.9 & 99.5 & 76.0 & 74.0 & 100 & 100.0 & 72.8 & 76.0\\
 &  &  & 0.3 & 35.4 & 99.1 & 92.7 & 92.0 & 99.7 & 86.7 & 87.2 & 100 & 100.0 & 90.0 & 92.8\\
 &  &  & 0.7 & 35.3 & 99.7 & 99.8 & 99.8 & 99.4 & 98.6 & 98.6 & 100 & 100.0 & 97.7 & 98.6\\
\addlinespace
 &  & 2:1 & 0.1 & 35.3 & 99.9 & 81.0 & 81.2 & 99.8 & 78.7 & 78.5 & 100 & 100.0 & 75.8 & 79.5\\
 &  &  & 0.3 & 35.3 & 99.5 & 96.7 & 95.2 & 99.5 & 93.5 & 92.9 & 100 & 100.0 & 94.3 & 96.0\\
 &  &  & 0.7 & 35.4 & 100.0 & 99.9 & 99.9 & 99.6 & 99.3 & 99.3 & 100 & 100.0 & 98.6 & 99.7\\
\bottomrule
\end{tabular}
\begin{tablenotes}
\item 
           Results of NN$_P$, HN$_P$, and CBN$_P$ are based on the population studies;
           results of NN$_O$, HN$_O$, and CBN$_O$ are based on the published studies;
           CN and CS indicate the Copas-N and Copas-Shi methods;
           HN$^\text{Prop}$ and CBN$^\text{Prop}$ indicate the proposed HN or CBN model based sensitivity analysis methods.
\end{tablenotes}
\end{threeparttable}
\end{table}

\begin{table}[!htbp]

\caption{\label{tab:set1-tau}Averages of estimates of $\tau^2$ among different models under the HN model based data-generating process.}
\centering
\begin{threeparttable}
\begin{tabular}[t]{rrrrrrrrrrrrrrr}
\toprule
$S$ & Patients & T:C & $\tau^2$ & $N$ & NN$_P$ & HN$_P$ & CBN$_P$ & NN$_O$ & HN$_O$ & CBN$_O$ & CN & CS & HN$^\text{Prop}$ & CBN$^\text{Prop}$\\
\midrule
15 & U[30, 60] & 1:1 & 0.1 & 10.6 & 0.04 & 0.21 & 0.12 & 0.05 & 0.23 & 0.14 & 0.73 & 0.00 & 0.32 & 0.21\\
 &  &  & 0.3 & 10.6 & 0.08 & 0.32 & 0.18 & 0.08 & 0.31 & 0.20 & 0.78 & 0.00 & 0.42 & 0.29\\
 &  &  & 0.7 & 10.6 & 0.25 & 0.56 & 0.38 & 0.21 & 0.50 & 0.36 & 0.88 & 0.01 & 0.60 & 0.45\\
\addlinespace
 &  & 2:1 & 0.1 & 10.6 & 0.06 & 0.18 & 0.06 & 0.06 & 0.18 & 0.07 & 0.74 & 0.01 & 0.27 & 0.13\\
 &  &  & 0.3 & 10.5 & 0.15 & 0.32 & 0.12 & 0.14 & 0.31 & 0.15 & 0.83 & 0.02 & 0.42 & 0.21\\
 &  &  & 0.7 & 10.5 & 0.33 & 0.56 & 0.28 & 0.29 & 0.51 & 0.30 & 0.91 & 0.09 & 0.61 & 0.38\\
\addlinespace
 & U[50, 200] & 1:1 & 0.1 & 10.7 & 0.02 & 0.20 & 0.16 & 0.02 & 0.19 & 0.18 & 0.64 & 0.00 & 0.30 & 0.25\\
 &  &  & 0.3 & 10.8 & 0.07 & 0.36 & 0.31 & 0.07 & 0.34 & 0.31 & 0.74 & 0.00 & 0.45 & 0.40\\
 &  &  & 0.7 & 10.8 & 0.18 & 0.56 & 0.51 & 0.16 & 0.51 & 0.47 & 0.85 & 0.00 & 0.61 & 0.56\\
\addlinespace
 &  & 2:1 & 0.1 & 10.9 & 0.04 & 0.18 & 0.14 & 0.03 & 0.19 & 0.15 & 0.70 & 0.00 & 0.26 & 0.21\\
 &  &  & 0.3 & 10.8 & 0.10 & 0.33 & 0.25 & 0.09 & 0.30 & 0.26 & 0.79 & 0.01 & 0.39 & 0.32\\
 &  &  & 0.7 & 10.8 & 0.25 & 0.58 & 0.49 & 0.21 & 0.53 & 0.46 & 0.89 & 0.05 & 0.60 & 0.53\\
\addlinespace
 & U[500, 700] & 1:1 & 0.1 & 10.4 & 0.02 & 0.22 & 0.22 & 0.03 & 0.24 & 0.23 & 0.63 & 0.00 & 0.34 & 0.31\\
 &  &  & 0.3 & 10.4 & 0.06 & 0.35 & 0.33 & 0.07 & 0.34 & 0.32 & 0.71 & 0.00 & 0.45 & 0.42\\
 &  &  & 0.7 & 10.4 & 0.15 & 0.54 & 0.53 & 0.14 & 0.49 & 0.49 & 0.81 & 0.00 & 0.59 & 0.56\\
\addlinespace
 &  & 2:1 & 0.1 & 10.3 & 0.03 & 0.18 & 0.17 & 0.03 & 0.20 & 0.19 & 0.69 & 0.00 & 0.29 & 0.26\\
 &  &  & 0.3 & 10.4 & 0.08 & 0.33 & 0.31 & 0.08 & 0.31 & 0.30 & 0.78 & 0.01 & 0.40 & 0.38\\
 &  &  & 0.7 & 10.4 & 0.20 & 0.58 & 0.56 & 0.18 & 0.51 & 0.50 & 0.89 & 0.03 & 0.60 & 0.57\\
\addlinespace
50 & U[30, 60] & 1:1 & 0.1 & 35.8 & 0.02 & 0.16 & 0.07 & 0.02 & 0.16 & 0.08 & 0.80 & 0.00 & 0.23 & 0.12\\
 &  &  & 0.3 & 35.9 & 0.07 & 0.32 & 0.15 & 0.06 & 0.29 & 0.16 & 0.89 & 0.00 & 0.36 & 0.22\\
 &  &  & 0.7 & 36.0 & 0.25 & 0.64 & 0.37 & 0.21 & 0.56 & 0.35 & 0.97 & 0.00 & 0.64 & 0.43\\
\addlinespace
 &  & 2:1 & 0.1 & 35.9 & 0.03 & 0.14 & 0.03 & 0.03 & 0.14 & 0.04 & 0.83 & 0.00 & 0.18 & 0.08\\
 &  &  & 0.3 & 35.9 & 0.11 & 0.30 & 0.08 & 0.10 & 0.27 & 0.10 & 0.93 & 0.01 & 0.34 & 0.15\\
 &  &  & 0.7 & 35.8 & 0.37 & 0.65 & 0.25 & 0.30 & 0.56 & 0.26 & 0.99 & 0.06 & 0.64 & 0.34\\
\addlinespace
 & U[50, 200] & 1:1 & 0.1 & 36.9 & 0.01 & 0.16 & 0.12 & 0.01 & 0.16 & 0.13 & 0.70 & 0.00 & 0.21 & 0.18\\
 &  &  & 0.3 & 37.1 & 0.04 & 0.31 & 0.25 & 0.04 & 0.28 & 0.24 & 0.81 & 0.00 & 0.34 & 0.30\\
 &  &  & 0.7 & 37.0 & 0.18 & 0.64 & 0.55 & 0.16 & 0.57 & 0.51 & 0.95 & 0.00 & 0.65 & 0.59\\
\addlinespace
 &  & 2:1 & 0.1 & 36.9 & 0.01 & 0.13 & 0.09 & 0.01 & 0.13 & 0.10 & 0.77 & 0.00 & 0.17 & 0.14\\
 &  &  & 0.3 & 37.0 & 0.07 & 0.31 & 0.22 & 0.06 & 0.27 & 0.21 & 0.90 & 0.00 & 0.33 & 0.27\\
 &  &  & 0.7 & 37.1 & 0.26 & 0.66 & 0.52 & 0.21 & 0.58 & 0.48 & 0.98 & 0.01 & 0.65 & 0.57\\
\addlinespace
 & U[500, 700] & 1:1 & 0.1 & 35.4 & 0.01 & 0.17 & 0.16 & 0.01 & 0.16 & 0.16 & 0.67 & 0.00 & 0.22 & 0.21\\
 &  &  & 0.3 & 35.4 & 0.04 & 0.31 & 0.30 & 0.04 & 0.30 & 0.28 & 0.78 & 0.00 & 0.36 & 0.34\\
 &  &  & 0.7 & 35.3 & 0.15 & 0.63 & 0.61 & 0.14 & 0.56 & 0.54 & 0.92 & 0.00 & 0.63 & 0.62\\
\addlinespace
 &  & 2:1 & 0.1 & 35.3 & 0.01 & 0.14 & 0.12 & 0.01 & 0.13 & 0.13 & 0.75 & 0.00 & 0.18 & 0.16\\
 &  &  & 0.3 & 35.3 & 0.05 & 0.29 & 0.28 & 0.05 & 0.27 & 0.25 & 0.87 & 0.00 & 0.32 & 0.31\\
 &  &  & 0.7 & 35.4 & 0.21 & 0.66 & 0.64 & 0.18 & 0.57 & 0.55 & 0.98 & 0.00 & 0.65 & 0.62\\
\bottomrule
\end{tabular}
\begin{tablenotes}
\item 
           Results of NN$_P$, HN$_P$, and CBN$_P$ are based on the population studies;
           results of NN$_O$, HN$_O$, and CBN$_O$ are based on the published studies;
           CN and CS indicate the Copas-N and Copas-Shi methods;
           HN$^\text{Prop}$ and CBN$^\text{Prop}$ indicate the proposed HN or CBN model based sensitivity analysis methods.
\end{tablenotes}
\end{threeparttable}
\end{table}

\clearpage
\subsection{Additional result 2: meta-analysis of odds ratios under the 2SBN model based data-generating process}

As mentioned in Sections 5.3, we presented convergence rates (Table \ref{tab:set2}) and average estimates of $\tau^2$ (Table \ref{tab:set2-tau}) when population data were simulated under the real-world data-generating process.

\begin{table}[!htbp]

\caption{\label{tab:set2}Convergence proportion (\%) of estimations among different models under the 2SBN model based data-generating process.}
\centering
\begin{threeparttable}
\begin{tabular}[t]{rrrrrrrrrrrrrrr}
\toprule
$S$ & Patients & T:C & $\tau^2$ & $N$ & NN$_P$ & HN$_P$ & CBN$_P$ & NN$_O$ & HN$_O$ & CBN$_O$ & CN & CS & HN$^\text{Prop}$ & CBN$^\text{Prop}$\\
\midrule
15 & U[30, 60] & 1:1 & 0.1 & 10.6 & 100.0 & 66.2 & 63.8 & 100.0 & 64.6 & 62.3 & 100 & 100.0 & 56.5 & 61.1\\
 &  &  & 0.3 & 10.6 & 100.0 & 69.7 & 70.1 & 99.9 & 69.7 & 64.6 & 100 & 100.0 & 63.6 & 72.2\\
 &  &  & 0.7 & 10.5 & 99.8 & 78.0 & 73.2 & 99.8 & 74.2 & 72.1 & 100 & 100.0 & 75.3 & 82.0\\
\addlinespace
 &  & 2:1 & 0.1 & 10.6 & 99.9 & 69.5 & 65.2 & 99.9 & 67.8 & 66.6 & 100 & 100.0 & 56.4 & 66.3\\
 &  &  & 0.3 & 10.6 & 100.0 & 70.5 & 69.5 & 99.9 & 69.8 & 65.7 & 100 & 100.0 & 66.5 & 73.9\\
 &  &  & 0.7 & 10.6 & 100.0 & 80.8 & 77.0 & 99.9 & 76.8 & 71.0 & 100 & 100.0 & 77.3 & 83.1\\
\addlinespace
 & U[50, 200] & 1:1 & 0.1 & 10.8 & 100.0 & 73.4 & 66.7 & 100.0 & 68.0 & 67.0 & 100 & 100.0 & 58.9 & 68.4\\
 &  &  & 0.3 & 10.8 & 99.9 & 80.3 & 78.8 & 99.8 & 77.7 & 76.5 & 100 & 100.0 & 71.2 & 77.2\\
 &  &  & 0.7 & 10.9 & 99.8 & 90.1 & 89.8 & 99.7 & 88.2 & 87.2 & 100 & 100.0 & 83.4 & 88.7\\
\addlinespace
 &  & 2:1 & 0.1 & 10.8 & 100.0 & 73.8 & 70.3 & 99.8 & 71.3 & 69.5 & 100 & 100.0 & 60.6 & 70.0\\
 &  &  & 0.3 & 10.9 & 99.4 & 85.5 & 80.4 & 99.5 & 79.5 & 77.1 & 100 & 100.0 & 73.0 & 80.8\\
 &  &  & 0.7 & 10.8 & 99.0 & 93.4 & 92.4 & 99.4 & 90.9 & 89.8 & 100 & 100.0 & 87.2 & 92.7\\
\addlinespace
 & U[500, 700] & 1:1 & 0.1 & 10.5 & 100.0 & 86.8 & 84.1 & 100.0 & 82.6 & 80.0 & 100 & 100.0 & 73.7 & 80.0\\
 &  &  & 0.3 & 10.5 & 100.0 & 98.3 & 97.7 & 100.0 & 94.3 & 93.7 & 100 & 100.0 & 88.7 & 94.3\\
 &  &  & 0.7 & 10.5 & 100.0 & 99.6 & 99.8 & 99.9 & 98.9 & 99.4 & 100 & 100.0 & 95.1 & 99.0\\
\addlinespace
 &  & 2:1 & 0.1 & 10.4 & 100.0 & 89.1 & 87.6 & 100.0 & 83.4 & 81.7 & 100 & 100.0 & 73.0 & 82.1\\
 &  &  & 0.3 & 10.4 & 99.9 & 98.4 & 98.3 & 99.9 & 94.3 & 95.8 & 100 & 100.0 & 89.8 & 94.2\\
 &  &  & 0.7 & 10.4 & 100.0 & 99.9 & 99.9 & 100.0 & 98.9 & 99.2 & 100 & 100.0 & 94.8 & 98.5\\
\addlinespace
50 & U[30, 60] & 1:1 & 0.1 & 36.0 & 100.0 & 69.7 & 65.1 & 99.9 & 70.5 & 66.2 & 100 & 100.0 & 63.1 & 68.8\\
 &  &  & 0.3 & 35.9 & 99.8 & 77.8 & 73.5 & 99.5 & 77.0 & 72.5 & 100 & 100.0 & 76.7 & 81.7\\
 &  &  & 0.7 & 35.9 & 98.7 & 90.4 & 86.5 & 98.7 & 87.4 & 83.9 & 100 & 100.0 & 84.4 & 89.0\\
\addlinespace
 &  & 2:1 & 0.1 & 36.1 & 100.0 & 71.1 & 68.5 & 100.0 & 69.6 & 67.1 & 100 & 100.0 & 66.5 & 71.0\\
 &  &  & 0.3 & 36.0 & 100.0 & 79.7 & 78.7 & 100.0 & 77.8 & 76.1 & 100 & 100.0 & 80.2 & 81.8\\
 &  &  & 0.7 & 35.9 & 99.8 & 92.1 & 90.8 & 99.5 & 90.3 & 87.1 & 100 & 100.0 & 90.9 & 92.8\\
\addlinespace
 & U[50, 200] & 1:1 & 0.1 & 37.2 & 99.0 & 76.4 & 77.0 & 99.5 & 76.0 & 72.7 & 100 & 100.0 & 73.0 & 75.2\\
 &  &  & 0.3 & 37.1 & 99.8 & 91.6 & 90.3 & 99.4 & 89.3 & 87.9 & 100 & 100.0 & 89.4 & 92.0\\
 &  &  & 0.7 & 37.0 & 99.6 & 99.1 & 98.9 & 99.6 & 98.1 & 98.3 & 100 & 100.0 & 97.6 & 98.7\\
\addlinespace
 &  & 2:1 & 0.1 & 37.0 & 99.6 & 79.3 & 76.1 & 99.8 & 77.9 & 75.8 & 100 & 100.0 & 75.6 & 79.2\\
 &  &  & 0.3 & 37.0 & 99.8 & 91.7 & 92.3 & 99.7 & 90.3 & 87.9 & 100 & 99.8 & 91.2 & 93.4\\
 &  &  & 0.7 & 37.1 & 99.8 & 99.4 & 99.5 & 99.8 & 99.3 & 99.0 & 100 & 100.0 & 98.4 & 99.3\\
\addlinespace
 & U[500, 700] & 1:1 & 0.1 & 35.2 & 100.0 & 97.4 & 97.4 & 99.7 & 94.4 & 93.1 & 100 & 100.0 & 90.5 & 93.4\\
 &  &  & 0.3 & 35.3 & 99.9 & 99.9 & 99.8 & 100.0 & 99.7 & 99.4 & 100 & 100.0 & 99.3 & 99.8\\
 &  &  & 0.7 & 35.3 & 100.0 & 100.0 & 100.0 & 100.0 & 100.0 & 100.0 & 100 & 100.0 & 99.4 & 100.0\\
\addlinespace
 &  & 2:1 & 0.1 & 35.3 & 99.9 & 97.4 & 98.1 & 100.0 & 94.9 & 93.6 & 100 & 100.0 & 92.0 & 94.4\\
 &  &  & 0.3 & 35.3 & 100.0 & 100.0 & 100.0 & 100.0 & 99.7 & 99.5 & 100 & 100.0 & 98.7 & 99.5\\
 &  &  & 0.7 & 35.4 & 100.0 & 100.0 & 100.0 & 100.0 & 100.0 & 100.0 & 100 & 100.0 & 98.9 & 100.0\\
\bottomrule
\end{tabular}
\begin{tablenotes}
\item 
           Results of NN$_P$, HN$_P$, and CBN$_P$ are based on the population studies;
           results of NN$_O$, HN$_O$, and CBN$_O$ are based on the published studies;
           CN and CS indicate the Copas-N and Copas-Shi methods;
           HN$^\text{Prop}$ and CBN$^\text{Prop}$ indicate the proposed HN or CBN model based sensitivity analysis methods.
\end{tablenotes}
\end{threeparttable}
\end{table}

\begin{table}[!htbp]

\caption{\label{tab:set2-tau}Averages of estimates of $\tau^2$ among different models under the 2SBN model based data-generating process.}
\centering
\begin{threeparttable}
\begin{tabular}[t]{rrrrrrrrrrrrrrr}
\toprule
$S$ & Patients & T:C & $\tau^2$ & $N$ & NN$_P$ & HN$_P$ & CBN$_P$ & NN$_O$ & HN$_O$ & CBN$_O$ & CN & CS & HN$^\text{Prop}$ & CBN$^\text{Prop}$\\
\midrule
15 & U[30, 60] & 1:1 & 0.1 & 10.6 & 0.00 & 0.30 & 0.26 & 0.00 & 0.29 & 0.25 & 0.90 & 0.00 & 0.45 & 0.39\\
 &  &  & 0.3 & 10.6 & 0.01 & 0.38 & 0.32 & 0.01 & 0.35 & 0.33 & 0.91 & 0.00 & 0.54 & 0.46\\
 &  &  & 0.7 & 10.5 & 0.03 & 0.49 & 0.46 & 0.04 & 0.47 & 0.41 & 0.92 & 0.01 & 0.61 & 0.54\\
\addlinespace
 &  & 2:1 & 0.1 & 10.6 & 0.00 & 0.31 & 0.27 & 0.01 & 0.31 & 0.26 & 0.85 & 0.00 & 0.49 & 0.40\\
 &  &  & 0.3 & 10.6 & 0.01 & 0.37 & 0.31 & 0.01 & 0.34 & 0.31 & 0.86 & 0.00 & 0.50 & 0.43\\
 &  &  & 0.7 & 10.6 & 0.04 & 0.50 & 0.46 & 0.05 & 0.48 & 0.45 & 0.88 & 0.02 & 0.63 & 0.55\\
\addlinespace
 & U[50, 200] & 1:1 & 0.1 & 10.8 & 0.02 & 0.20 & 0.18 & 0.03 & 0.21 & 0.18 & 0.92 & 0.01 & 0.30 & 0.25\\
 &  &  & 0.3 & 10.8 & 0.06 & 0.35 & 0.31 & 0.07 & 0.32 & 0.29 & 0.92 & 0.02 & 0.44 & 0.38\\
 &  &  & 0.7 & 10.9 & 0.17 & 0.53 & 0.48 & 0.18 & 0.48 & 0.44 & 0.94 & 0.07 & 0.58 & 0.52\\
\addlinespace
 &  & 2:1 & 0.1 & 10.8 & 0.02 & 0.19 & 0.15 & 0.03 & 0.19 & 0.15 & 0.80 & 0.01 & 0.28 & 0.22\\
 &  &  & 0.3 & 10.9 & 0.07 & 0.33 & 0.30 & 0.08 & 0.32 & 0.29 & 0.83 & 0.04 & 0.42 & 0.36\\
 &  &  & 0.7 & 10.8 & 0.22 & 0.58 & 0.53 & 0.23 & 0.52 & 0.48 & 0.89 & 0.10 & 0.61 & 0.55\\
\addlinespace
 & U[500, 700] & 1:1 & 0.1 & 10.5 & 0.05 & 0.10 & 0.09 & 0.05 & 0.09 & 0.08 & 0.68 & 0.02 & 0.13 & 0.12\\
 &  &  & 0.3 & 10.5 & 0.19 & 0.28 & 0.25 & 0.17 & 0.24 & 0.22 & 0.76 & 0.08 & 0.30 & 0.27\\
 &  &  & 0.7 & 10.5 & 0.44 & 0.60 & 0.55 & 0.40 & 0.52 & 0.47 & 0.86 & 0.26 & 0.59 & 0.54\\
\addlinespace
 &  & 2:1 & 0.1 & 10.4 & 0.06 & 0.10 & 0.09 & 0.06 & 0.09 & 0.08 & 0.47 & 0.03 & 0.13 & 0.11\\
 &  &  & 0.3 & 10.4 & 0.20 & 0.28 & 0.26 & 0.18 & 0.24 & 0.22 & 0.60 & 0.11 & 0.30 & 0.27\\
 &  &  & 0.7 & 10.4 & 0.46 & 0.59 & 0.55 & 0.41 & 0.51 & 0.47 & 0.78 & 0.30 & 0.58 & 0.54\\
\addlinespace
50 & U[30, 60] & 1:1 & 0.1 & 36.0 & 0.00 & 0.28 & 0.22 & 0.00 & 0.26 & 0.22 & 0.98 & 0.00 & 0.39 & 0.31\\
 &  &  & 0.3 & 35.9 & 0.00 & 0.38 & 0.33 & 0.00 & 0.38 & 0.33 & 0.98 & 0.00 & 0.51 & 0.42\\
 &  &  & 0.7 & 35.9 & 0.01 & 0.55 & 0.49 & 0.02 & 0.52 & 0.45 & 0.98 & 0.00 & 0.64 & 0.53\\
\addlinespace
 &  & 2:1 & 0.1 & 36.1 & 0.00 & 0.28 & 0.21 & 0.00 & 0.30 & 0.23 & 0.92 & 0.00 & 0.40 & 0.31\\
 &  &  & 0.3 & 36.0 & 0.00 & 0.39 & 0.31 & 0.00 & 0.38 & 0.30 & 0.93 & 0.00 & 0.48 & 0.39\\
 &  &  & 0.7 & 35.9 & 0.01 & 0.61 & 0.52 & 0.02 & 0.57 & 0.49 & 0.95 & 0.01 & 0.66 & 0.57\\
\addlinespace
 & U[50, 200] & 1:1 & 0.1 & 37.2 & 0.01 & 0.16 & 0.12 & 0.01 & 0.14 & 0.12 & 0.99 & 0.00 & 0.20 & 0.16\\
 &  &  & 0.3 & 37.1 & 0.04 & 0.30 & 0.26 & 0.05 & 0.28 & 0.24 & 0.99 & 0.01 & 0.35 & 0.29\\
 &  &  & 0.7 & 37.0 & 0.18 & 0.62 & 0.55 & 0.20 & 0.57 & 0.49 & 1.00 & 0.08 & 0.64 & 0.56\\
\addlinespace
 &  & 2:1 & 0.1 & 37.0 & 0.01 & 0.16 & 0.13 & 0.01 & 0.16 & 0.12 & 0.89 & 0.00 & 0.21 & 0.16\\
 &  &  & 0.3 & 37.0 & 0.04 & 0.31 & 0.25 & 0.05 & 0.28 & 0.23 & 0.92 & 0.02 & 0.34 & 0.28\\
 &  &  & 0.7 & 37.1 & 0.21 & 0.65 & 0.58 & 0.23 & 0.59 & 0.52 & 0.97 & 0.12 & 0.66 & 0.58\\
\addlinespace
 & U[500, 700] & 1:1 & 0.1 & 35.2 & 0.05 & 0.09 & 0.08 & 0.05 & 0.08 & 0.07 & 0.83 & 0.02 & 0.11 & 0.09\\
 &  &  & 0.3 & 35.3 & 0.20 & 0.29 & 0.27 & 0.18 & 0.25 & 0.23 & 0.91 & 0.11 & 0.30 & 0.27\\
 &  &  & 0.7 & 35.3 & 0.47 & 0.67 & 0.61 & 0.43 & 0.58 & 0.52 & 0.97 & 0.32 & 0.65 & 0.59\\
\addlinespace
 &  & 2:1 & 0.1 & 35.3 & 0.06 & 0.09 & 0.08 & 0.06 & 0.08 & 0.08 & 0.57 & 0.03 & 0.11 & 0.09\\
 &  &  & 0.3 & 35.3 & 0.21 & 0.28 & 0.26 & 0.19 & 0.24 & 0.22 & 0.70 & 0.13 & 0.29 & 0.26\\
 &  &  & 0.7 & 35.4 & 0.50 & 0.67 & 0.61 & 0.46 & 0.58 & 0.53 & 0.90 & 0.36 & 0.67 & 0.61\\
\bottomrule
\end{tabular}
\begin{tablenotes}
\item 
           Results of NN$_P$, HN$_P$, and CBN$_P$ are based on the population studies;
           results of NN$_O$, HN$_O$, and CBN$_O$ are based on the published studies;
           CN and CS indicate the Copas-N and Copas-Shi methods;
           HN$^\text{Prop}$ and CBN$^\text{Prop}$ indicate the proposed HN or CBN model based sensitivity analysis methods.
\end{tablenotes}
\end{threeparttable}
\end{table}

\clearpage
\subsection{Additional result 3: meta-analysis of proportions under the 1SBN model data-generating process}

As mentioned in Sections 5.4, we presented convergence rates (Table \ref{tab:set3}) and average estimates of $\tau^2$ (Table \ref{tab:set3-tau}) when population data were simulated under the SGBN model based data-generating process.

\begin{table}[!htbp]

\caption{\label{tab:set3}Convergence proportion (\%) of estimations among different models under the SGBN model based data-generating process.}
\centering
\begin{threeparttable}
\begin{tabular}[t]{rrrrrrrrrrr}
\toprule
$S$ & Patients & $\tau^2$ & $N$ & NN$_P$ & 1SBN$_P$ & NN$_O$ & 1SBN$_O$ & CN & CS & 1SBN$^{\text{prop}}$\\
\midrule
15 & U[15, 30] & 0.1 & 10.6 & 100.0 & 78.4 & 99.9 & 76.3 & 100 & 100 & 74.6\\
 &  & 0.3 & 10.6 & 99.6 & 92.8 & 99.8 & 89.5 & 100 & 100 & 89.5\\
 &  & 0.7 & 10.6 & 100.0 & 97.9 & 100.0 & 96.4 & 100 & 100 & 96.9\\
\addlinespace
 & U[25, 100] & 0.1 & 10.8 & 100.0 & 91.1 & 99.9 & 90.4 & 100 & 100 & 88.1\\
 &  & 0.3 & 10.8 & 100.0 & 99.2 & 99.9 & 97.4 & 100 & 100 & 97.2\\
 &  & 0.7 & 10.8 & 100.0 & 100.0 & 100.0 & 99.1 & 100 & 100 & 99.5\\
\addlinespace
50 & U[15, 30] & 0.1 & 36.0 & 99.6 & 89.4 & 99.8 & 85.3 & 100 & 100 & 86.7\\
 &  & 0.3 & 36.1 & 99.9 & 99.6 & 99.7 & 98.6 & 100 & 100 & 98.7\\
 &  & 0.7 & 36.2 & 100.0 & 100.0 & 100.0 & 99.9 & 100 & 100 & 99.8\\
\addlinespace
 & U[25, 100] & 0.1 & 37.0 & 100.0 & 99.4 & 100.0 & 97.8 & 100 & 100 & 98.1\\
 &  & 0.3 & 37.1 & 100.0 & 100.0 & 100.0 & 100.0 & 100 & 100 & 99.9\\
 &  & 0.7 & 37.0 & 100.0 & 100.0 & 100.0 & 100.0 & 100 & 100 & 100.0\\
\bottomrule
\end{tabular}
\begin{tablenotes}
\item 
           NN$_P$ and BN$_P$ are the estimates based on the population studies;
           NN$_O$ and BN$_O$ are the estimates based on the published studies;
           CN and CS are the Copas-N and Copas-Shi methods;
           1SBN$^{\text{prop}}$ are the proposed 1SBN model based sensitivity analysis methods.
\end{tablenotes}
\end{threeparttable}
\end{table}

\begin{table}[!htbp]

\caption{\label{tab:set3-tau}Averages of estimates of $\tau^2$ among different models under the SGBN model based data-generating process.}
\centering
\begin{threeparttable}
\begin{tabular}[t]{rrrrrrrrrrr}
\toprule
$S$ & Patients & $\tau^2$ & $N$ & NN$_P$ & 1SBN$_P$ & NN$_O$ & 1SBN$_O$ & CN & CS & 1SBN$^{\text{prop}}$\\
\midrule
15 & U[15, 30] & 0.1 & 10.6 & 0.04 & 0.14 & 0.04 & 0.13 & 0.52 & 0.00 & 0.18\\
 &  & 0.3 & 10.6 & 0.12 & 0.30 & 0.12 & 0.28 & 0.67 & 0.00 & 0.35\\
 &  & 0.7 & 10.6 & 0.32 & 0.59 & 0.30 & 0.53 & 0.83 & 0.01 & 0.59\\
\addlinespace
 & U[25, 100] & 0.1 & 10.8 & 0.06 & 0.10 & 0.06 & 0.09 & 0.26 & 0.00 & 0.12\\
 &  & 0.3 & 10.8 & 0.20 & 0.28 & 0.19 & 0.25 & 0.44 & 0.05 & 0.28\\
 &  & 0.7 & 10.8 & 0.47 & 0.61 & 0.45 & 0.54 & 0.73 & 0.22 & 0.59\\
\addlinespace
50 & U[15, 30] & 0.1 & 36.0 & 0.03 & 0.11 & 0.03 & 0.11 & 0.55 & 0.00 & 0.13\\
 &  & 0.3 & 36.1 & 0.12 & 0.29 & 0.11 & 0.25 & 0.72 & 0.00 & 0.30\\
 &  & 0.7 & 36.2 & 0.35 & 0.66 & 0.34 & 0.59 & 0.93 & 0.00 & 0.65\\
\addlinespace
 & U[25, 100] & 0.1 & 37.0 & 0.06 & 0.10 & 0.06 & 0.08 & 0.26 & 0.00 & 0.10\\
 &  & 0.3 & 37.1 & 0.22 & 0.29 & 0.21 & 0.26 & 0.47 & 0.07 & 0.29\\
 &  & 0.7 & 37.0 & 0.52 & 0.68 & 0.50 & 0.60 & 0.84 & 0.30 & 0.67\\
\bottomrule
\end{tabular}
\begin{tablenotes}
\item 
           NN$_P$ and BN$_P$ are the estimates based on the population studies;
           NN$_O$ and BN$_O$ are the estimates based on the published studies;
           CN and CS are the Copas-N and Copas-Shi methods;
           1SBN$^{\text{prop}}$ are the proposed 1SBN model based sensitivity analysis methods.
\end{tablenotes}
\end{threeparttable}
\end{table}

\clearpage
\section{Estimations in Application}\label{sec:app}

\subsection{Example 1: rare-event meta-analysis of odds ratios}
As mentioned in Section 6.1, we presented the data of example 1 in Table \ref{tab:eg1}.
The estimates of parameters were presented in Table \ref{tab:tab1-1}-\ref{tab:tab1-4}.

\begin{table}[!htbp]
\caption{Data of catheter-related bloodstream infection}
\centering
\label{tab:eg1}
\begin{tabular}{@{}rrrrr@{}}
\toprule
      & \multicolumn{2}{r}{Standard catheter}           & \multicolumn{2}{r}{AIT catheter}             \\ \cmidrule(l){2-5} 
Study & CRBSI ($\Ya$) & Patients ($\Na$) & CRBSI ($\Yb$) & Patients ($\Nb$) \\ \midrule
1     & 3                  & 117                & 0                  & 116                \\
2     & 3                  & 35                 & 1                  & 44                 \\
3     & 9                  & 195                & 2                  & 208                \\
4     & 7                  & 136                & 0                  & 130                \\
5     & 6                  & 157                & 5                  & 151                \\
6     & 4                  & 139                & 1                  & 98                 \\
7     & 3                  & 177                & 1                  & 174                \\
8     & 2                  & 39                 & 1                  & 74                 \\
9     & 19                 & 103                & 1                  & 97                 \\
10    & 2                  & 122                & 1                  & 113                \\
11    & 7                  & 64                 & 0                  & 66                 \\
12    & 1                  & 58                 & 0                  & 70                 \\
13    & 5                  & 175                & 3                  & 188                \\
14    & 11                 & 180                & 6                  & 187                \\
15    & 0                  & 105                & 0                  & 118                \\
16    & 1                  & 262                & 0                  & 252                \\
17    & 3                  & 362                & 1                  & 345                \\
18    & 1                  & 69                 & 4                  & 64                 \\ \bottomrule
\end{tabular}%
\end{table}

\begin{table}[!htbp]

\caption{\label{tab:tab1-1}Example 1: summary of the estimations of different sensitivity analysis methods}
\centering
\begin{threeparttable}
\begin{tabular}[t]{rrrrrrrr}
\toprule
\multicolumn{1}{c}{ } & \multicolumn{1}{c}{} & \multicolumn{3}{c}{The proposed HN model based method} & \multicolumn{3}{c}{The proposed CBN model based method} \\
\cmidrule(l{3pt}r{3pt}){3-5} \cmidrule(l{3pt}r{3pt}){6-8}
$(\Pmin, \Pmax)$ & $M$ & $\theta$ (95\% CI) & $\tau$ & $\rho$ & $\theta$ (95\% CI) & $\tau$ & $\rho$\\
\midrule
(0.99, 0.999) & 0 & -1.352 (-2.047, -0.657) & 0.833 & -0.121 & -1.301 (-1.972, -0.631) & 0.775 & -0.119\\
(0.90, 0.999) & 1 & -1.345 (-2.060, -0.629) & 0.834 & -0.154 & -1.295 (-1.985, -0.605) & 0.776 & -0.152\\
(0.80, 0.999) & 1 & -1.337 (-2.073, -0.601) & 0.834 & -0.169 & -1.288 (-1.998, -0.578) & 0.776 & -0.168\\
(0.70, 0.999) & 2 & -1.330 (-2.086, -0.573) & 0.835 & -0.179 & -1.281 (-2.010, -0.552) & 0.777 & -0.179\\
(0.60, 0.999) & 3 & -1.321 (-2.098, -0.545) & 0.835 & -0.186 & -1.273 (-2.022, -0.525) & 0.777 & -0.186\\
(0.50, 0.999) & 4 & -1.312 (-2.110, -0.515) & 0.835 & -0.190 & -1.265 (-2.034, -0.496) & 0.777 & -0.190\\
(0.40, 0.999) & 6 & -1.302 (-2.123, -0.482) & 0.835 & -0.193 & -1.256 (-2.046, -0.465) & 0.777 & -0.193\\
(0.30, 0.999) & 9 & -1.291 (-2.137, -0.446) & 0.835 & -0.194 & -1.245 (-2.060, -0.430) & 0.776 & -0.194\\
(0.20, 0.999) & 14 & -1.277 (-2.152, -0.402) & 0.834 & -0.193 & -1.232 (-2.076, -0.388) & 0.776 & -0.193\\
(0.10, 0.999) & 26 & -1.258 (-2.174, -0.342) & 0.832 & -0.187 & -1.214 (-2.097, -0.331) & 0.774 & -0.188\\
\bottomrule
\end{tabular}
\begin{tablenotes}
\item $M$ indicates the number of potentially unpublished studies; 
           CI indicates the confidence interval.
\end{tablenotes}
\end{threeparttable}
\end{table}

\begin{table}[!htbp]

\caption{\label{tab:tab1-2}Example 1: summary of the estimations of different sensitivity analysis methods}
\centering
\begin{threeparttable}
\begin{tabular}[t]{rrrrrrrrr}
\toprule
\multicolumn{1}{c}{ } & \multicolumn{4}{c}{The Copas-N method (only0)} & \multicolumn{4}{c}{The Copas-N method (all)} \\
\cmidrule(l{3pt}r{3pt}){2-5} \cmidrule(l{3pt}r{3pt}){6-9}
$(\Pmin, \Pmax)$ & $M_1$ & $\theta$ (95\% CI) & $\tau$ & $\rho$ & $M_2$ & $\theta$ (95\% CI) & $\tau$ & $\rho$\\
\midrule
(0.99, 0.999) & 0 & -1.177 (-1.669, -0.685) & 1.052 & -0.051 & 0 & -1.077 (-1.540, -0.615) & 0.987 & -0.077\\
(0.90, 0.999) & 1 & -1.172 (-1.683, -0.662) & 1.052 & -0.071 & 1 & -1.071 (-1.554, -0.589) & 0.988 & -0.102\\
(0.80, 0.999) & 2 & -1.168 (-1.697, -0.638) & 1.053 & -0.079 & 2 & -1.065 (-1.567, -0.563) & 0.988 & -0.112\\
(0.70, 0.999) & 3 & -1.163 (-1.711, -0.614) & 1.053 & -0.084 & 3 & -1.059 (-1.580, -0.538) & 0.989 & -0.117\\
(0.60, 0.999) & 4 & -1.157 (-1.725, -0.589) & 1.053 & -0.087 & 4 & -1.052 (-1.592, -0.512) & 0.989 & -0.120\\
(0.50, 0.999) & 6 & -1.152 (-1.740, -0.563) & 1.053 & -0.089 & 5 & -1.045 (-1.604, -0.486) & 0.989 & -0.121\\
(0.40, 0.999) & 9 & -1.145 (-1.755, -0.536) & 1.053 & -0.090 & 8 & -1.038 (-1.617, -0.459) & 0.989 & -0.121\\
(0.30, 0.999) & 13 & -1.138 (-1.772, -0.505) & 1.053 & -0.090 & 11 & -1.029 (-1.630, -0.429) & 0.989 & -0.120\\
(0.20, 0.999) & 21 & -1.130 (-1.790, -0.470) & 1.053 & -0.088 & 18 & -1.019 (-1.644, -0.395) & 0.989 & -0.117\\
(0.10, 0.999) & 40 & -1.119 (-1.814, -0.423) & 1.053 & -0.085 & 35 & -1.006 (-1.663, -0.349) & 0.989 & -0.111\\
\bottomrule
\end{tabular}
\begin{tablenotes}
\item $M_1$ and $M_2$ indicate the number of potentially unpublished studies; 
  only 0 indicates continuity correction for only studies with 0 cells;
  all indicates continuity correction for all the studies; 
           CI indicates the confidence interval.
\end{tablenotes}
\end{threeparttable}
\end{table}

\begin{table}[!htbp]

\caption{\label{tab:tab1-3}Example 1: summary of the estimations of different sensitivity analysis methods}
\centering
\begin{threeparttable}
\begin{tabular}[t]{rrrrrrrrr}
\toprule
\multicolumn{1}{c}{ } & \multicolumn{4}{c}{The Copas-Shi method (only0)} & \multicolumn{4}{c}{The Copas-Shi method (all)} \\
\cmidrule(l{3pt}r{3pt}){2-5} \cmidrule(l{3pt}r{3pt}){6-9}
$(\Pmin, \Pmax)$ & $M_1$ & $\theta$ (95\% CI) & $\tau$ & $\rho$ & $M_2$ & $\theta$ (95\% CI) & $\tau$ & $\rho$\\
\midrule
(0.99, 0.999) & 0 & -0.950 (-1.412, -0.488) & 0.001 & -0.369 & 0 & -0.857 (-1.280, -0.434) & 0.001 & -0.409\\
(0.90, 0.999) & 1 & -0.918 (-1.389, -0.446) & 0.001 & -0.516 & 1 & -0.830 (-1.259, -0.401) & 0.001 & -0.565\\
(0.80, 0.999) & 2 & -0.882 (-1.363, -0.400) & 0.001 & -0.574 & 2 & -0.801 (-1.236, -0.365) & 0.001 & -0.622\\
(0.70, 0.999) & 3 & -0.845 (-1.338, -0.351) & 0.001 & -0.601 & 3 & -0.771 (-1.214, -0.328) & 0.001 & -0.646\\
(0.60, 0.999) & 4 & -0.808 (-1.320, -0.296) & 0.001 & -0.603 & 4 & -0.741 (-1.194, -0.288) & 0.001 & -0.647\\
(0.50, 0.999) & 6 & -0.776 (-1.311, -0.240) & 0.001 & -0.582 & 5 & -0.715 (-1.183, -0.247) & 0.001 & -0.623\\
(0.40, 0.999) & 9 & -0.751 (-1.311, -0.191) & 0.001 & -0.540 & 8 & -0.695 (-1.182, -0.208) & 0.001 & -0.577\\
(0.30, 0.999) & 13 & -0.734 (-1.311, -0.157) & 0.001 & -0.484 & 11 & -0.682 (-1.185, -0.179) & 0.001 & -0.513\\
(0.20, 0.999) & 21 & -0.722 (-1.310, -0.134) & 0.001 & -0.420 & 18 & -0.674 (-1.187, -0.161) & 0.001 & -0.441\\
(0.10, 0.999) & 40 & -0.711 (-1.308, -0.114) & 0.001 & -0.347 & 35 & -0.667 (-1.187, -0.148) & 0.001 & -0.360\\
\bottomrule
\end{tabular}
\begin{tablenotes}
\item $M_1$ and $M_2$ indicate the number of potentially unpublished studies; 
  only 0 indicates continuity correction for only studies with 0 cells;
  all indicates continuity correction for all the studies; 
           CI indicates the confidence interval.
\end{tablenotes}
\end{threeparttable}
\end{table}

\begin{table}[!htbp]

\caption{\label{tab:tab1-4}Example 1: summary of the estimations of different sensitivity analysis methods}
\centering
\begin{threeparttable}
\begin{tabular}[t]{rrrrrrr}
\toprule
\multicolumn{1}{c}{ } & \multicolumn{1}{c}{} & \multicolumn{2}{c}{The t-statistic and HN model based method (only0)} & \multicolumn{2}{c}{The t-statistic and HN model based method (all)} \\
\cmidrule(l{3pt}r{3pt}){3-4} \cmidrule(l{3pt}r{3pt}){5-6}
$p$ & $M_1$ & $\theta$ (95\% CI) & $\tau$ & $M_2$ & $\theta$ (95\% CI) & $\tau$\\
\midrule
1.00 & 0 & -1.353 (-2.041, -0.665) & 0.833 & 0 & -1.353 (-2.041, -0.665) & 0.833\\
0.90 & 2 & -1.204 (-1.939, -0.470) & 0.970 & 2 & -1.209 (-1.945, -0.473) & 0.963\\
0.80 & 4 & -1.033 (-1.897, -0.168) & 1.079 & 4 & -1.042 (-1.870, -0.215) & 1.070\\
0.70 & 8 & -0.842 (-1.891, 0.208) & 1.173 & 8 & -0.856 (-1.908, 0.195) & 1.162\\
0.60 & 12 & -0.631 (-1.996, 0.734) & 1.255 & 12 & -0.652 (-1.980, 0.677) & 1.243\\
0.50 & 18 & -0.404 (-2.227, 1.419) & 1.323 & 18 & -0.429 (-2.154, 1.296) & 1.312\\
0.40 & 27 & -0.171 (-2.690, 2.349) & 1.370 & 27 & -0.192 (-2.461, 2.077) & 1.365\\
0.30 & 42 & 0.027 (-3.554, 3.608) & 1.377 & 42 & 0.043 (-2.952, 3.039) & 1.394\\
0.20 & 72 & 0.093 (-4.498, 4.684) & 1.313 & 72 & 0.238 (-3.658, 4.134) & 1.382\\
0.10 & 162 & 0.017 (-4.927, 4.961) & 1.207 & 162 & 0.302 (-4.339, 4.944) & 1.307\\
\bottomrule
\end{tabular}
\begin{tablenotes}
\item $M$ indicates the number of potentially unpublished studies; 
  only 0 indicates continuity correction for only studies with 0 cells;
  all indicates continuity correction for all the studies; 
           CI indicates the confidence interval.
\end{tablenotes}
\end{threeparttable}
\end{table}

\clearpage
\subsection{Example 2: rare-event meta-analysis of odds ratios}

As mentioned in Section 6.2, we presented the data of example 2 in Table \ref{tab:eg2}.
The estimates of parameters were shown in Table \ref{tab:tab2-1}-\ref{tab:tab2-4}.

\begin{table}[!htbp]
\caption{Data of catheter-related bloodstream infection}
\centering
\label{tab:eg2}
\begin{tabular}{@{}rrrrr@{}}
\toprule
      & \multicolumn{2}{r}{Magnesium group}           & \multicolumn{2}{r}{Control group}             \\ \cmidrule(l){2-5} 
Study & Deaths  ($\Ya$) & Patients ($\Na$) & Deaths ($\Yb$) & Patients ($\Nb$) \\ \midrule
1 & 1 & 40 & 2 & 36\\
2 & 9 & 135 & 23 & 135\\
3 & 2 & 200 & 7 & 200\\
4 & 1 & 48 & 1 & 46\\
5 & 10 & 150 & 8 & 148\\
6 & 1 & 59 & 9 & 56\\
7 & 1 & 25 & 3 & 23\\
8 & 0 & 22 & 1 & 21\\
9 & 6 & 76 & 11 & 75\\
10 & 1 & 27 & 7 & 27\\
11 & 2 & 89 & 12 & 80\\
12 & 5 & 23 & 13 & 33\\
13 & 4 & 130 & 8 & 122\\
14 & 90 & 1159 & 118 & 1157\\
15 & 4 & 107 & 17 & 108\\
16 & 2216 & 29011 & 2103 & 29039\\
\bottomrule
\end{tabular}%
\end{table}

\begin{table}[!htbp]

\caption{\label{tab:tab2-1}Example 2: summary of the estimations of different sensitivity analysis methods}
\centering
\begin{threeparttable}
\begin{tabular}[t]{rrrrrrrr}
\toprule
\multicolumn{1}{c}{ } & \multicolumn{1}{c}{} & \multicolumn{3}{c}{The proposed HN model based method} & \multicolumn{3}{c}{The proposed CBN model based method} \\
\cmidrule(l{3pt}r{3pt}){3-5} \cmidrule(l{3pt}r{3pt}){6-8}
$(\Pmin, \Pmax)$ & $M$ & $\theta$ (95\% CI) & $\tau$ & $\rho$ & $\theta$ (95\% CI) & $\tau$ & $\rho$\\
\midrule
(0.99, 0.999) & 0 & -0.841 (-1.295, -0.388) & 0.569 & -0.999 & -0.750 (-1.174, -0.325) & 0.510 & -0.990\\
(0.90, 0.999) & 1 & -0.796 (-1.227, -0.364) & 0.592 & -0.999 & -0.711 (-1.114, -0.308) & 0.530 & -0.990\\
(0.80, 0.999) & 3 & -0.733 (-1.132, -0.334) & 0.608 & -0.999 & -0.654 (-1.025, -0.283) & 0.547 & -0.990\\
(0.70, 0.999) & 6 & -0.658 (-1.022, -0.294) & 0.625 & -0.999 & -0.586 (-0.926, -0.247) & 0.561 & -0.990\\
(0.60, 0.999) & 9 & -0.573 (-0.909, -0.237) & 0.641 & -0.999 & -0.511 (-0.827, -0.194) & 0.574 & -0.990\\
(0.50, 0.999) & 13 & -0.477 (-0.799, -0.154) & 0.656 & -0.999 & -0.427 (-0.735, -0.120) & 0.584 & -0.990\\
(0.40, 0.999) & 19 & -0.368 (-0.700, -0.037) & 0.667 & -0.999 & -0.334 (-0.655, -0.014) & 0.590 & -0.990\\
(0.30, 0.999) & 30 & -0.243 (-0.615, 0.130) & 0.674 & -0.999 & -0.228 (-0.590, 0.133) & 0.592 & -0.990\\
(0.20, 0.999) & 50 & -0.089 (-0.545, 0.366) & 0.675 & -0.999 & -0.101 (-0.541, 0.338) & 0.587 & -0.990\\
(0.10, 0.999) & 107 & 0.118 (-0.490, 0.726) & 0.662 & -0.999 & 0.065 (-0.511, 0.640) & 0.566 & -0.990\\
\bottomrule
\end{tabular}
\begin{tablenotes}
\item $M$ indicates the number of potentially unpublished studies; 
           CI indicates the confidence interval.
\end{tablenotes}
\end{threeparttable}
\end{table}
\begin{table}[!htbp]

\caption{\label{tab:tab2-2}Example 2: summary of the estimations of different sensitivity analysis methods}
\centering
\begin{threeparttable}
\begin{tabular}[t]{rrrrrrrrr}
\toprule
\multicolumn{1}{c}{ } & \multicolumn{4}{c}{The Copas-N method (only0)} & \multicolumn{4}{c}{The Copas-N method (all)} \\
\cmidrule(l{3pt}r{3pt}){2-5} \cmidrule(l{3pt}r{3pt}){6-9}
$(\Pmin, \Pmax)$ & $M_1$ & $\theta$ (95\% CI) & $\tau$ & $\rho$ & $M_2$ & $\theta$ (95\% CI) & $\tau$ & $\rho$\\
\midrule
(0.99, 0.999) & 0 & -1.013 (-1.387, -0.640) & 0.756 & -0.990 & 0 & -0.905 (-1.232, -0.577) & 0.661 & -0.990\\
(0.90, 0.999) & 1 & -0.870 (-1.190, -0.551) & 0.865 & -0.990 & 1 & -0.768 (-1.052, -0.484) & 0.768 & -0.990\\
(0.80, 0.999) & 3 & -0.808 (-1.206, -0.410) & 0.871 & -0.871 & 3 & -0.731 (-1.089, -0.373) & 0.755 & -0.843\\
(0.70, 0.999) & 6 & -0.704 (-1.130, -0.277) & 0.905 & -0.863 & 6 & -0.645 (-1.042, -0.249) & 0.781 & -0.834\\
(0.60, 0.999) & 9 & -0.589 (-1.081, -0.098) & 0.935 & -0.860 & 9 & -0.555 (-1.023, -0.086) & 0.802 & -0.828\\
(0.50, 0.999) & 13 & -0.466 (-1.072, 0.140) & 0.959 & -0.856 & 13 & -0.460 (-1.034, 0.114) & 0.817 & -0.819\\
(0.40, 0.999) & 19 & -0.337 (-1.123, 0.449) & 0.975 & -0.846 & 19 & -0.362 (-1.068, 0.345) & 0.825 & -0.806\\
(0.30, 0.999) & 29 & -0.210 (-1.238, 0.817) & 0.977 & -0.825 & 29 & -0.261 (-1.114, 0.592) & 0.826 & -0.787\\
(0.20, 0.999) & 49 & -0.093 (-1.359, 1.172) & 0.959 & -0.790 & 48 & -0.154 (-1.162, 0.855) & 0.816 & -0.760\\
(0.10, 0.999) & 107 & 0.018 (-1.427, 1.463) & 0.918 & -0.732 & 104 & -0.037 (-1.208, 1.135) & 0.791 & -0.715\\
\bottomrule
\end{tabular}
\begin{tablenotes}
\item $M_1$ and $M_2$ indicate the number of potentially unpublished studies; 
  only 0 indicates continuity correction for only studies with 0 cells;
  all indicates continuity correction for all the studies; 
           CI indicates the confidence interval.
\end{tablenotes}
\end{threeparttable}
\end{table}
\begin{table}[!htbp]

\caption{\label{tab:tab2-3}Example 2: summary of the estimations of different sensitivity analysis methods}
\centering
\begin{threeparttable}
\begin{tabular}[t]{rrrrrrrrr}
\toprule
\multicolumn{1}{c}{ } & \multicolumn{4}{c}{The Copas-Shi method (only0)} & \multicolumn{4}{c}{The Copas-Shi method (all)} \\
\cmidrule(l{3pt}r{3pt}){2-5} \cmidrule(l{3pt}r{3pt}){6-9}
$(\Pmin, \Pmax)$ & $M_1$ & $\theta$ (95\% CI) & $\tau$ & $\rho$ & $M_2$ & $\theta$ (95\% CI) & $\tau$ & $\rho$\\
\midrule
(0.99, 0.999) & 0 & -0.737 (-1.187, -0.288) & 0.500 & -0.561 & 0 & -0.683 (-1.092, -0.273) & 0.459 & -0.541\\
(0.90, 0.999) & 1 & -0.647 (-1.103, -0.190) & 0.459 & -0.828 & 1 & -0.603 (-1.019, -0.186) & 0.424 & -0.817\\
(0.80, 0.999) & 3 & -0.534 (-0.977, -0.090) & 0.409 & -0.931 & 3 & -0.499 (-0.905, -0.092) & 0.378 & -0.928\\
(0.70, 0.999) & 6 & -0.405 (-0.859, 0.050) & 0.356 & -0.990 & 6 & -0.377 (-0.796, 0.041) & 0.328 & -0.990\\
(0.60, 0.999) & 9 & -0.256 (-0.716, 0.205) & 0.281 & -0.990 & 9 & -0.238 (-0.656, 0.180) & 0.259 & -0.990\\
(0.50, 0.999) & 13 & -0.132 (-0.476, 0.213) & 0.191 & -0.957 & 13 & -0.121 (-0.440, 0.197) & 0.182 & -0.962\\
(0.40, 0.999) & 19 & 0.045 (-0.015, 0.106) & 0.001 & -0.936 & 19 & 0.046 (-0.015, 0.106) & 0.001 & -0.939\\
(0.30, 0.999) & 29 & 0.046 (-0.016, 0.108) & 0.001 & -0.875 & 29 & 0.046 (-0.016, 0.108) & 0.001 & -0.876\\
(0.20, 0.999) & 49 & 0.046 (-0.015, 0.108) & 0.001 & -0.796 & 48 & 0.046 (-0.015, 0.108) & 0.001 & -0.796\\
(0.10, 0.999) & 107 & 0.046 (-0.015, 0.108) & 0.001 & -0.697 & 104 & 0.046 (-0.015, 0.108) & 0.001 & -0.695\\
\bottomrule
\end{tabular}
\begin{tablenotes}
\item $M_1$ and $M_2$ indicate the number of potentially unpublished studies; 
  only 0 indicates continuity correction for only studies with 0 cells;
  all indicates continuity correction for all the studies; 
           CI indicates the confidence interval.
\end{tablenotes}
\end{threeparttable}
\end{table}

\begin{table}[!htbp]

\caption{\label{tab:tab2-4}Example 2: summary of the estimations of different sensitivity analysis methods}
\centering
\begin{threeparttable}
\begin{tabular}[t]{rrrrrrr}
\toprule
\multicolumn{1}{c}{ } & \multicolumn{1}{c}{} & \multicolumn{2}{c}{The t-statistic and HN model based method (only0)} & \multicolumn{2}{c}{The t-statistic and HN model based method (all)} \\
\cmidrule(l{3pt}r{3pt}){3-4} \cmidrule(l{3pt}r{3pt}){5-6}
$p$ & $M_1$ & $\theta$ (95\% CI) & $\tau$ & $M_2$ & $\theta$ (95\% CI) & $\tau$\\
\midrule
1.00 & 0 & -0.951 (-1.392, -0.510) & 0.457 & 0 & -0.951 (-1.392, -0.510) & 0.457\\
0.90 & 2 & -0.945 (-1.388, -0.502) & 0.463 & 2 & -0.945 (-1.394, -0.496) & 0.462\\
0.80 & 4 & -0.917 (-1.223, -0.611) & 0.548 & 4 & -0.918 (-1.254, -0.582) & 0.537\\
0.70 & 7 & -1.041 (-1.465, -0.616) & 0.535 & 7 & -1.033 (-1.440, -0.626) & 0.521\\
0.60 & 11 & -1.127 (NaN, NaN) & 0.510 & 11 & -1.113 (NaN, NaN) & 0.495\\
0.50 & 16 & -1.185 (NaN, NaN) & 0.479 & 16 & -1.168 (NaN, NaN) & 0.464\\
0.40 & 24 & -1.224 (-2.633, 0.186) & 0.444 & 24 & -1.204 (-1.603, -0.806) & 0.430\\
0.30 & 37 & -1.249 (-1.474, -1.024) & 0.408 & 37 & -1.229 (-1.424, -1.034) & 0.395\\
0.20 & 64 & -1.265 (-1.424, -1.106) & 0.368 & 64 & -1.246 (-1.394, -1.098) & 0.357\\
0.10 & 144 & -1.276 (-1.416, -1.136) & 0.319 & 144 & -1.259 (-1.394, -1.124) & 0.311\\
\bottomrule
\end{tabular}
\begin{tablenotes}
\item $M$ indicates the number of potentially unpublished studies; 
  only 0 indicates continuity correction for only studies with 0 cells;
  all indicates continuity correction for all the studies; 
           CI indicates the confidence interval.
\end{tablenotes}
\end{threeparttable}
\end{table}

\clearpage
\subsection{Example 3: rare-event meta-analysis of proportions}

The estimates of parameters were shown in Table \ref{tab:tab3-1}-\ref{tab:tab3-4}.

\begin{table}[!htbp]

\caption{\label{tab:tab3-1}Example 3: summary of the estimations of different sensitivity analysis methods}
\centering
\begin{threeparttable}
\begin{tabular}[t]{rrrrr}
\toprule
\multicolumn{1}{c}{ } & \multicolumn{1}{c}{} & \multicolumn{3}{c}{The proposed 1SBN model based method} \\
\cmidrule(l{3pt}r{3pt}){3-5}
$(\Pmin, \Pmax)$ & $M$ & $\theta$ (95\% CI) & $\tau$ & $\rho$\\
\midrule
(0.99, 0.999) & 0 & -4.818 (-5.515, -4.122) & 0.912 & 0.990\\
(0.90, 0.999) & 1 & -4.850 (-5.554, -4.146) & 0.929 & 0.990\\
(0.80, 0.999) & 1 & -4.885 (-5.599, -4.170) & 0.945 & 0.990\\
(0.70, 0.999) & 2 & -4.923 (-5.650, -4.195) & 0.960 & 0.990\\
(0.60, 0.999) & 3 & -4.965 (-5.709, -4.221) & 0.974 & 0.990\\
(0.50, 0.999) & 4 & -5.013 (-5.779, -4.247) & 0.986 & 0.990\\
(0.40, 0.999) & 6 & -5.069 (-5.863, -4.275) & 0.996 & 0.990\\
(0.30, 0.999) & 9 & -5.136 (-5.968, -4.304) & 1.001 & 0.990\\
(0.20, 0.999) & 14 & -5.096 (-6.206, -3.987) & 0.950 & 0.729\\
(0.10, 0.999) & 27 & -5.088 (-6.124, -4.051) & 0.921 & 0.558\\
\bottomrule
\end{tabular}
\begin{tablenotes}
\item $M$ indicates the number of potentially unpublished studies; 
           CI indicates the confidence interval.
\end{tablenotes}
\end{threeparttable}
\end{table}
\begin{table}[!htbp]

\caption{\label{tab:tab3-2}Example 3: summary of the estimations of different sensitivity analysis methods}
\centering
\begin{threeparttable}
\begin{tabular}[t]{rrrrrrrrr}
\toprule
\multicolumn{1}{c}{ } & \multicolumn{4}{c}{The Copas-N method (only0)} & \multicolumn{4}{c}{The Copas-N method (all)} \\
\cmidrule(l{3pt}r{3pt}){2-5} \cmidrule(l{3pt}r{3pt}){6-9}
$(\Pmin, \Pmax)$ & $M_1$ & $\theta$ (95\% CI) & $\tau$ & $\rho$ & $M_2$ & $\theta$ (95\% CI) & $\tau$ & $\rho$\\
\midrule
(0.99, 0.999) & 0 & -4.652 (-5.069, -4.235) & 0.896 & 0.985 & 0 & -4.460 (-4.894, -4.026) & 0.934 & 0.990\\
(0.90, 0.999) & 1 & -4.652 (-5.069, -4.235) & 0.896 & 0.990 & 1 & -4.460 (-4.894, -4.026) & 0.934 & 0.990\\
(0.80, 0.999) & 2 & -4.652 (-5.069, -4.235) & 0.896 & 0.990 & 2 & -4.460 (-4.894, -4.027) & 0.934 & 0.990\\
(0.70, 0.999) & 4 & -4.652 (-5.067, -4.238) & 0.896 & 0.990 & 3 & -4.474 (-4.863, -4.085) & 0.940 & 0.990\\
(0.60, 0.999) & 6 & -4.671 (-5.029, -4.313) & 0.901 & 0.990 & 5 & -4.549 (-4.857, -4.241) & 0.965 & 0.990\\
(0.50, 0.999) & 9 & -4.751 (-5.028, -4.474) & 0.918 & 0.990 & 8 & -4.670 (-4.932, -4.407) & 0.995 & 0.990\\
(0.40, 0.999) & 13 & -4.876 (-5.111, -4.641) & 0.938 & 0.990 & 11 & -4.819 (-5.058, -4.581) & 1.028 & 0.990\\
(0.30, 0.999) & 20 & -5.036 (-5.255, -4.816) & 0.960 & 0.990 & 17 & -5.004 (-5.241, -4.768) & 1.066 & 0.990\\
(0.20, 0.999) & 34 & -5.245 (-5.481, -5.010) & 0.987 & 0.990 & 29 & -5.054 (-5.417, -4.691) & 1.104 & 0.927\\
(0.10, 0.999) & 71 & -5.424 (-5.743, -5.105) & 1.017 & 0.957 & 62 & -5.021 (-5.781, -4.262) & 1.022 & 0.775\\
\bottomrule
\end{tabular}
\begin{tablenotes}
\item $M_1$ and $M_2$ indicate the number of potentially unpublished studies; 
  only 0 indicates continuity correction for only studies with 0 cells;
  all indicates continuity correction for all the studies; 
           CI indicates the confidence interval.
\end{tablenotes}
\end{threeparttable}
\end{table}
\begin{table}[!htbp]

\caption{\label{tab:tab3-3}Example 3: summary of the estimations of different sensitivity analysis methods}
\centering
\begin{threeparttable}
\begin{tabular}[t]{rrrrrrrrr}
\toprule
\multicolumn{1}{c}{ } & \multicolumn{4}{c}{The Copas-Shi method (only0)} & \multicolumn{4}{c}{The Copas-Shi method (all)} \\
\cmidrule(l{3pt}r{3pt}){2-5} \cmidrule(l{3pt}r{3pt}){6-9}
$(\Pmin, \Pmax)$ & $M_1$ & $\theta$ (95\% CI) & $\tau$ & $\rho$ & $M_2$ & $\theta$ (95\% CI) & $\tau$ & $\rho$\\
\midrule
(0.99, 0.999) & 0 & -4.229 (-4.786, -3.672) & 0.576 & -0.481 & 0 & -4.023 (-4.516, -3.531) & 0.510 & -0.556\\
(0.90, 0.999) & 1 & -4.156 (-4.714, -3.598) & 0.533 & -0.713 & 1 & -3.965 (-4.459, -3.471) & 0.473 & -0.762\\
(0.80, 0.999) & 2 & -4.072 (-4.624, -3.520) & 0.481 & -0.813 & 2 & -3.906 (-4.392, -3.419) & 0.435 & -0.837\\
(0.70, 0.999) & 4 & -3.981 (-4.524, -3.438) & 0.419 & -0.879 & 3 & -3.845 (-4.322, -3.368) & 0.393 & -0.889\\
(0.60, 0.999) & 6 & -3.877 (-4.415, -3.340) & 0.337 & -0.930 & 5 & -3.778 (-4.246, -3.310) & 0.344 & -0.933\\
(0.50, 0.999) & 9 & -3.723 (-4.055, -3.390) & 0.001 & -0.977 & 8 & -3.636 (-3.907, -3.365) & 0.001 & -0.990\\
(0.40, 0.999) & 13 & -3.684 (-4.035, -3.333) & 0.001 & -0.990 & 11 & -3.601 (-3.877, -3.326) & 0.001 & -0.990\\
(0.30, 0.999) & 20 & -3.454 (-3.781, -3.126) & 0.001 & -0.990 & 17 & -3.442 (-3.757, -3.126) & 0.001 & -0.990\\
(0.20, 0.999) & 34 & -3.483 (-3.913, -3.053) & 0.001 & -0.829 & 29 & -3.472 (-3.847, -3.097) & 0.001 & -0.817\\
(0.10, 0.999) & 71 & -3.479 (-3.918, -3.040) & 0.001 & -0.701 & 62 & -3.461 (-3.845, -3.077) & 0.001 & -0.699\\
\bottomrule
\end{tabular}
\begin{tablenotes}
\item $M_1$ and $M_2$ indicate the number of potentially unpublished studies; 
  only 0 indicates continuity correction for only studies with 0 cells;
  all indicates continuity correction for all the studies; 
           CI indicates the confidence interval.
\end{tablenotes}
\end{threeparttable}
\end{table}
\begin{table}[!htbp]

\caption{\label{tab:tab3-4}Example 3: summary of the estimations of different sensitivity analysis methods}
\centering
\begin{threeparttable}
\begin{tabular}[t]{rrrrrr}
\toprule
\multicolumn{1}{c}{ } & \multicolumn{1}{c}{} & \multicolumn{2}{c}{The t-statistic and HN model based method (only0)} & \multicolumn{2}{c}{The t-statistic and HN model based method (all)} \\
\cmidrule(l{3pt}r{3pt}){3-4} \cmidrule(l{3pt}r{3pt}){5-6}
$p$ & $M$ & $\theta$ (95\% CI) & $\tau$ & $\theta$ (95\% CI) & $\tau$\\
\midrule
1.00 & 0 & -4.812 (-5.508, -4.116) & 0.908 & -4.812 (-5.508, -4.116) & 0.908\\
0.90 & 2 & -4.672 (-5.078, -4.266) & 0.929 & -4.557 (-4.787, -4.326) & 0.961\\
0.80 & 4 & -4.418 (-5.149, -3.687) & 0.863 & -4.513 (-5.124, -3.903) & 0.889\\
0.70 & 8 & -4.431 (-4.996, -3.866) & 0.834 & -4.438 (-5.068, -3.807) & 0.859\\
0.60 & 12 & -4.357 (-4.927, -3.786) & 0.798 & -4.369 (-5.032, -3.705) & 0.833\\
0.50 & 18 & -4.285 (-4.869, -3.701) & 0.764 & -4.304 (-5.004, -3.603) & 0.808\\
0.40 & 27 & -4.214 (-4.820, -3.608) & 0.729 & -4.240 (-4.980, -3.500) & 0.783\\
0.30 & 42 & -4.139 (-4.767, -3.512) & 0.691 & -4.173 (-4.953, -3.392) & 0.756\\
0.20 & 72 & -4.055 (-4.706, -3.404) & 0.646 & -4.097 (-4.926, -3.268) & 0.724\\
0.10 & 162 & -3.943 (-4.618, -3.269) & 0.582 & -3.995 (-4.876, -3.114) & 0.678\\
\bottomrule
\end{tabular}
\begin{tablenotes}
\item $M$ indicates the number of potentially unpublished studies; 
  only 0 indicates continuity correction for only studies with 0 cells;
  all indicates continuity correction for all the studies; 
           CI indicates the confidence interval.
\end{tablenotes}
\end{threeparttable}
\end{table}

\clearpage
\subsection{Example 4: rare-event meta-analysis of proportions}

As mentioned in Section 6.2, we presented the data of example 4 in Table \ref{tab:eg4}.

\begin{table}[!htbp]

\caption{\label{tab:eg4}Data of Example 4}
\centering
\begin{tabular}[t]{rrrr}
\toprule
Study & Improved ($\Ya$) & Patients ($\Nb$) & Not improved ($\Yb$)\\
\midrule
1 & 16 & 17 & 1\\
2 & 10 & 12 & 2\\
3 & 4 & 8 & 4\\
4 & 43 & 58 & 15\\
5 & 10 & 10 & 0\\
6 & 25 & 42 & 17\\
7 & 13 & 14 & 1\\
8 & 12 & 12 & 0\\
9 & 22 & 41 & 19\\
10 & 4 & 5 & 1\\
11 & 5 & 6 & 1\\
12 & 18 & 23 & 5\\
13 & 58 & 68 & 10\\
14 & 6 & 10 & 4\\
\bottomrule
\end{tabular}
\end{table}

\begin{table}[!htbp]

\caption{\label{tab:tab4-1}Example 4: summary of the estimations of different sensitivity analysis methods}
\centering
\begin{threeparttable}
\begin{tabular}[t]{rrrrr}
\toprule
\multicolumn{1}{c}{ } & \multicolumn{1}{c}{} & \multicolumn{3}{c}{The proposed 1SBN model based method} \\
\cmidrule(l{3pt}r{3pt}){3-5}
$(\Pmin, \Pmax)$ & $M$ & $\theta$ (95\% CI) & $\tau$ & $\rho$\\
\midrule
(0.99, 0.999) & 0 & -1.374 (-1.941, -0.808) & 0.774 & -0.990\\
(0.90, 0.999) & 1 & -1.338 (-1.893, -0.782) & 0.805 & -0.990\\
(0.80, 0.999) & 1 & -1.291 (-1.826, -0.755) & 0.813 & -0.990\\
(0.70, 0.999) & 2 & -1.265 (-1.830, -0.700) & 0.805 & -0.871\\
(0.60, 0.999) & 3 & -1.245 (-1.831, -0.659) & 0.798 & -0.776\\
(0.50, 0.999) & 5 & -1.224 (-1.827, -0.621) & 0.790 & -0.709\\
(0.40, 0.999) & 7 & -1.203 (-1.823, -0.583) & 0.783 & -0.658\\
(0.30, 0.999) & 10 & -1.180 (-1.817, -0.542) & 0.776 & -0.614\\
(0.20, 0.999) & 16 & -1.153 (-1.811, -0.495) & 0.767 & -0.570\\
(0.10, 0.999) & 32 & -1.119 (-1.803, -0.434) & 0.754 & -0.516\\
\bottomrule
\end{tabular}
\begin{tablenotes}
\item $M$ indicates the number of potentially unpublished studies; 
           CI indicates the confidence interval.
\end{tablenotes}
\end{threeparttable}
\end{table}
\begin{table}[!htbp]

\caption{\label{tab:tab4-2}Example 4: summary of the estimations of different sensitivity analysis methods}
\centering
\begin{threeparttable}
\begin{tabular}[t]{rrrrrrrrr}
\toprule
\multicolumn{1}{c}{ } & \multicolumn{4}{c}{The Copas-N method (only0)} & \multicolumn{4}{c}{The Copas-N method (all)} \\
\cmidrule(l{3pt}r{3pt}){2-5} \cmidrule(l{3pt}r{3pt}){6-9}
$(\Pmin, \Pmax)$ & $M_1$ & $\theta$ (95\% CI) & $\tau$ & $\rho$ & $M_2$ & $\theta$ (95\% CI) & $\tau$ & $\rho$\\
\midrule
(0.99, 0.999) & 0 & -1.512 (-2.055, -0.969) & 0.998 & -0.990 & 0 & -1.393 (-1.911, -0.876) & 0.947 & -0.990\\
(0.90, 0.999) & 1 & -1.424 (-1.885, -0.964) & 1.059 & -0.990 & 1 & -1.326 (-1.767, -0.885) & 0.994 & -0.990\\
(0.80, 0.999) & 1 & -1.361 (-2.087, -0.635) & 1.081 & -0.836 & 1 & -1.146 (-1.562, -0.729) & 1.110 & -0.990\\
(0.70, 0.999) & 2 & -1.406 (-2.100, -0.711) & 1.026 & -0.483 & 2 & -1.315 (-2.017, -0.612) & 0.963 & -0.375\\
(0.60, 0.999) & 4 & -1.392 (-2.083, -0.702) & 1.021 & -0.417 & 3 & -1.307 (-1.993, -0.621) & 0.960 & -0.316\\
(0.50, 0.999) & 5 & -1.376 (-2.080, -0.672) & 1.018 & -0.379 & 5 & -1.295 (-1.990, -0.601) & 0.958 & -0.287\\
(0.40, 0.999) & 8 & -1.358 (-2.080, -0.636) & 1.015 & -0.351 & 7 & -1.282 (-1.991, -0.573) & 0.956 & -0.266\\
(0.30, 0.999) & 12 & -1.338 (-2.081, -0.594) & 1.011 & -0.326 & 10 & -1.267 (-1.993, -0.541) & 0.955 & -0.249\\
(0.20, 0.999) & 19 & -1.314 (-2.082, -0.546) & 1.007 & -0.302 & 15 & -1.249 (-1.995, -0.502) & 0.952 & -0.232\\
(0.10, 0.999) & 39 & -1.283 (-2.082, -0.484) & 1.002 & -0.273 & 31 & -1.224 (-1.997, -0.451) & 0.949 & -0.212\\
\bottomrule
\end{tabular}
\begin{tablenotes}
\item $M_1$ and $M_2$ indicate the number of potentially unpublished studies; 
  only 0 indicates continuity correction for only studies with 0 cells;
  all indicates continuity correction for all the studies; 
           CI indicates the confidence interval.
\end{tablenotes}
\end{threeparttable}
\end{table}
\begin{table}[!htbp]

\caption{\label{tab:tab4-3}Example 4: summary of the estimations of different sensitivity analysis methods}
\centering
\begin{threeparttable}
\begin{tabular}[t]{rrrrrrrrr}
\toprule
\multicolumn{1}{c}{ } & \multicolumn{4}{c}{The Copas-Shi method (only0)} & \multicolumn{4}{c}{The Copas-Shi method (all)} \\
\cmidrule(l{3pt}r{3pt}){2-5} \cmidrule(l{3pt}r{3pt}){6-9}
$(\Pmin, \Pmax)$ & $M_1$ & $\theta$ (95\% CI) & $\tau$ & $\rho$ & $M_2$ & $\theta$ (95\% CI) & $\tau$ & $\rho$\\
\midrule
(0.99, 0.999) & 0 & -1.102 (-1.577, -0.626) & 0.535 & -0.990 & 0 & -1.042 (-1.474, -0.611) & 0.495 & -0.990\\
(0.90, 0.999) & 1 & -1.037 (-1.492, -0.583) & 0.495 & -0.990 & 1 & -0.994 (-1.413, -0.576) & 0.466 & -0.990\\
(0.80, 0.999) & 1 & -0.992 (-1.434, -0.549) & 0.475 & -0.990 & 1 & -0.958 (-1.369, -0.548) & 0.451 & -0.990\\
(0.70, 0.999) & 2 & -0.953 (-1.386, -0.519) & 0.465 & -0.990 & 2 & -0.926 (-1.330, -0.522) & 0.442 & -0.990\\
(0.60, 0.999) & 4 & -0.916 (-1.339, -0.492) & 0.459 & -0.985 & 3 & -0.893 (-1.289, -0.497) & 0.436 & -0.990\\
(0.50, 0.999) & 5 & -0.869 (-1.278, -0.459) & 0.451 & -0.990 & 5 & -0.855 (-1.241, -0.468) & 0.430 & -0.990\\
(0.40, 0.999) & 8 & -0.804 (-1.194, -0.414) & 0.441 & -0.990 & 7 & -0.803 (-1.176, -0.431) & 0.423 & -0.990\\
(0.30, 0.999) & 12 & -0.693 (-1.060, -0.326) & 0.435 & -0.990 & 10 & -0.718 (-1.068, -0.368) & 0.410 & -0.990\\
(0.20, 0.999) & 19 & -0.736 (-1.209, -0.264) & 0.439 & -0.810 & 15 & -0.729 (-1.176, -0.282) & 0.423 & -0.829\\
(0.10, 0.999) & 39 & -0.742 (-1.212, -0.273) & 0.439 & -0.671 & 31 & -0.733 (-1.177, -0.289) & 0.422 & -0.685\\
\bottomrule
\end{tabular}
\begin{tablenotes}
\item $M_1$ and $M_2$ indicate the number of potentially unpublished studies; 
  only 0 indicates continuity correction for only studies with 0 cells;
  all indicates continuity correction for all the studies; 
           CI indicates the confidence interval.
\end{tablenotes}
\end{threeparttable}
\end{table}
\begin{table}[!htbp]

\caption{\label{tab:tab4-4}Example 4: summary of the estimations of different sensitivity analysis methods}
\centering
\begin{threeparttable}
\begin{tabular}[t]{rrrrrr}
\toprule
\multicolumn{1}{c}{ } & \multicolumn{1}{c}{} & \multicolumn{2}{c}{The t-statistic and HN model based method (only0)} & \multicolumn{2}{c}{The t-statistic and HN model based method (all)} \\
\cmidrule(l{3pt}r{3pt}){3-4} \cmidrule(l{3pt}r{3pt}){5-6}
$p$ & $M$ & $\theta$ (95\% CI) & $\tau$ & $\theta$ (95\% CI) & $\tau$\\
\midrule
1.00 & 0 & -1.377 (-1.942, -0.811) & 0.768 & -1.377 (-1.942, -0.811) & 0.768\\
0.90 & 2 & -1.446 (-2.013, -0.879) & 0.776 & -1.439 (-2.002, -0.876) & 0.770\\
0.80 & 4 & -1.500 (-2.079, -0.922) & 0.771 & -1.490 (-2.058, -0.922) & 0.763\\
0.70 & 6 & -0.721 (-1.110, -0.332) & 1.417 & -0.721 (-1.109, -0.332) & 1.418\\
0.60 & 9 & -0.374 (-0.693, -0.055) & 1.631 & -0.373 (-0.693, -0.054) & 1.632\\
0.50 & 14 & 0.052 (-0.280, 0.383) & 1.853 & 0.052 (-0.279, 0.383) & 1.853\\
0.40 & 21 & -1.670 (-2.302, -1.038) & 0.718 & -1.651 (-2.261, -1.041) & 0.707\\
0.30 & 33 & -1.710 (-2.360, -1.060) & 0.700 & -1.689 (-2.314, -1.065) & 0.687\\
0.20 & 56 & -1.752 (-2.424, -1.081) & 0.677 & -1.731 (-2.372, -1.089) & 0.664\\
0.10 & 126 & -1.803 (-2.507, -1.100) & 0.648 & -1.781 (-2.448, -1.113) & 0.633\\
\bottomrule
\end{tabular}
\begin{tablenotes}
\item $M$ indicates the number of potentially unpublished studies; 
  only 0 indicates continuity correction for only studies with 0 cells;
  all indicates continuity correction for all the studies; 
           CI indicates the confidence interval.
\end{tablenotes}
\end{threeparttable}
\end{table}

\clearpage

\subsection{Computational time}

As mentioned in Section xx of the main text, we summarized the computational time of the proposed methods and the t-statistic based method by setting 10 scenarios of sensitivity parameters.
The computations were implemented by R (version 4.0.3) on system of Intel Xeon Platinum 8368.
Parallel computations were conducted using R function \texttt{mclapply} by setting 10 cores.

\begin{table}[!htbp]
\caption{Computational time (seconds) of the proposed methods and t-statistic based method}
\label{tab:time}
\centering
\begin{threeparttable}
\begin{tabular}{@{}rrrrr@{}}
\toprule
 & Example 1 & Example 2 & Example 3 & Example 4 \\ \midrule
\begin{tabular}[c]{@{}r@{}}Proposed methods \\ (both the HN and BN models based)\end{tabular} & 17 & 34 & 4 & 3 \\ \addlinespace
\begin{tabular}[c]{@{}r@{}}The t-statistic based method\\ (only the HN model based)\end{tabular} & 97 & 3071 & 419 & 84 \\ \bottomrule
\end{tabular}
\end{threeparttable}
\end{table}


\end{document}